\documentclass[useAMS,usenatbib,fleqn]{mn2e}

\usepackage{times}
\usepackage{graphicx}
\usepackage{amssymb}
\usepackage{amsmath}
\usepackage{subfig}
\usepackage{url}
\usepackage{xspace}
\usepackage{gensymb}
\usepackage{enumerate}

\newcommand{\Msol}{M_{\odot}}

\newcommand{\xm}{\langle x \rangle_\mathrm{m} }

\newcommand{\ud}{\mathrm{d} }
\newcommand\ion[2]{#1$\,${\scshape{#2}}}%
\newcommand\hi{\ion{H}{i}\xspace}

\newcommand{\deltah}{\delta_{\rho_{\mathrm{H}}}}
\newcommand{\deltahi}{\delta_{\rho_{\mathrm{HI}}}}
\newcommand{\ctwo}{\textsc{C}$^2$\textsc{-ray}\xspace}
\newcommand{\cthree}{\textsc{CubeP}$^3$\textsc{M}\xspace}

\title[21-cm redshift space distortions with LOFAR]{Probing reionization with LOFAR using 21-cm redshift space distortions}
\author[H. Jensen et al.]{Hannes Jensen$^{1}$\thanks{hannes.jensen@astro.su.se},
Kanan K.\ Datta$^{1}$, Garrelt Mellema$^{1}$, Emma Chapman$^{2}$, \newauthor
Filipe B.\ Abdalla$^{2}$, Ilian T.\ Iliev$^{3}$, Yi Mao$^4$, Mario G. Santos$^{5,6}$, Paul R.\ Shapiro$^{7}$, \newauthor
Saleem Zaroubi$^8$, G. Bernardi$^{8,9}$, M. A. Brentjens$^{10}$, A. G. de Bruyn$^{8,10}$, B. Ciardi$^{11}$, \newauthor
G. J. A. Harker$^{12}$, V. Jeli\'{c}$^{8,10}$, S. Kazemi$^8$, L. V. E. Koopmans$^8$, P. Labropoulos$^{10}$,  \newauthor
O. Martinez$^8$,  A. R. Offringa$^{8,13}$, V. N. Pandey$^{10}$, J. Schaye$^{14}$, R. M. Thomas$^8$, \newauthor 
V. Veligatla$^8$, H. Vedantham$^8$, S. Yatawatta$^{10}$ \\
$^{1}$Dept.\ of Astronomy and Oskar Klein Centre, Stockholm University, AlbaNova, SE-10691 Stockholm, Sweden\\
$^{2}$Department of Physics and Astronomy, University College London, Gower Street, London WC1E 6BT\\
$^{3}$Astronomy Centre, Department of Physics \& Astronomy, Pevensey II Building, University of Sussex, Falmer, Brighton BN1 9QH\\
$^4$UPMC Univ Paris 06, CNRS, Institut Lagrange de Paris, Institut d’Astrophysique de Paris, \\
UMR7095, 98 bis, Boulevard Arago, F-75014, Paris, France\\
$^5$CENTRA, Instituto Superior Tecnico, Technical University of Lisbon, Lisboa 1049-001, Portugal \\
$^{6}$Department of Physics, University of the Western Cape, Bellville 7535, South Africa \\
$^{7}$Department of Astronomy, University of Texas, Austin, TX 78712-1083, USA\\
$^8$University of Groningen, Kapteyn Astronomical Institute, PO Box 800, 9700 AV Groningen, The Netherlands\\
$^9$Harvard-Smithsonian Center for Astrophysics, 60 Garden Street, Cambridge, MA 02138, USA\\
$^{10}$ASTRON, PO Box 2, 7990 AA Dwingeloo, The Netherlands\\
$^{11}$Max-Planck Institute for Astrophysics, Karl-Schwarzschild-Strasse 1, 85748 Garching bei M\"unchen, Germany\\
$^{12}$Center for Astrophysics and Space Astronomy, University of Colorado, 389 UCB, Boulder, Colorado 80309-0389, USA\\
$^{13}$Mount Stromlo Observatory, RSAA, Cotter Road, Weston Creek, ACT 2611, Australia\\
$^{14}$Leiden Observatory, Leiden University, PO Box 9513, 2300 RA Leiden, The Netherlands\\
}%\\
\begin{document}

\date{\today}

\pagerange{\pageref{firstpage}--\pageref{lastpage}} \pubyear{2012}

\maketitle

\label{firstpage}

\begin{abstract}
One of the most promising ways to study the epoch of reionization (EoR) is through radio observations of the redshifted 21-cm line emission from neutral hydrogen. These observations are complicated by the fact that the mapping of redshifts to line-of-sight positions is distorted by the peculiar velocities of the gas. Such distortions can be a source of error if they are not properly understood, but they also encode information about cosmology and astrophysics. We study the effects of redshift space distortions on the power spectrum of 21-cm radiation from the EoR using large scale $N$-body and radiative transfer simulations. We quantify the anisotropy introduced in the 21-cm power spectrum by redshift space distortions and show how it evolves as reionization progresses and how it relates to the underlying physics. We go on to study the effects of redshift space distortions on LOFAR observations, taking instrument noise and foreground subtraction into account. We find that LOFAR should be able to directly observe the power spectrum anisotropy due to redshift space distortions at spatial scales around $k \sim 0.1$~Mpc$^{-1}$ after $\gtrsim$ 1000 hours of integration time. At larger scales, sample errors become a limiting factor, while at smaller scales detector noise and foregrounds make the extraction of the signal problematic. Finally, we show how the astrophysical information contained in the evolution of the anisotropy of the 21-cm power spectrum can be extracted from LOFAR observations, and how it can be used to distinguish between different reionization scenarios.
\end{abstract}

%%%%%%%%%%%%%%%%%%%%%%%%%%%%%%%%%%%%%%%%%%%%%%%%

\section{Introduction}
\label{:sec:introduction}
During the past century, ever deeper observations have been mapping out the structure and history of the Universe, all the way back to the emission of the cosmic microwave background radiation (CMB) at the time of last scattering (e.g.\ \citealt{smoot1992,percival2001,abazajian2003}). However, there is an important gap in the observations, between the emission of the CMB at $z\approx1100$ and redshifts $z\approx7$. During this largely uncharted time period, the Universe was transformed from a featureless expanse of neutral gas into stars and galaxies surrounded by an ionized plasma.

Observing this time period, known as the epoch of reionization (EoR), is one of the current frontiers in observational cosmology. Indirect constraints can be obtained from observables such as quasar spectra \citep{fan2006,mortlock2011}, the CMB polarisation \citep{komatsu2011,larson2011} and temperature measurements \citep{theuns2002,raskutti2012}. Meanwhile, space-based galaxy surveys are pushing the redshift limits ever higher (e.g.\ \citealt{ellis2013}). Still, however, the details about the timing of the EoR and the sources that drove it are largely unknown. 

One of the most promising prospects for studying the process of reionization in detail is observations of the highly redshifted 21-cm emission originating from the neutral hydrogen in the intergalactic medium (IGM) as it is being ionized by the first sources of light. A number of radio interferometers---including LOFAR \citep{vanhaarlem2013}, PAPER \citep{parsons2010}, 21CMA \citep{wang2013}, GMRT \citep{pen2008,ali2008} and MWA \citep{tingay2012}---are just beginning observations hopefully leading up to the detection of the redshifted 21-cm radiation from the EoR in the near future. In this paper, we focus on LOFAR, which is a multi-purpose radio interferometer operated by a Dutch-led European collaboration, with one of its key science projects being the EoR. In late 2012, the LOFAR EoR project started observations of several fields, with the hope of making the first-ever detection of the 21-cm signal from the EoR within the near future \citep{debruyn2011,yatawatta2013}.

It has been shown that after long observation times, instruments such as LOFAR and MWA may be able to reach noise levels low enough to directly image the largest structures during the EoR \citep{datta2012a,zaroubi2012,malloy2012,chapman2013}. However, for this first generation of telescopes, most focus will be on statistics of the 21-cm signal---most notably the power spectrum, which contains a wealth of information about the physics of reionization (e.g.\ \citealt{pritchard2008}). 

Since the observable is an emission line, it is possible to translate observations at a specific frequency to a redshift, which in turn can be mapped to a position along the line-of-sight to produce three-dimensional data. However, the signal will be distorted by the peculiar velocities of the gas. Since matter tends to move toward higher-density regions, peculiar velocities introduce non-random distortions to the 21-cm signal, changing the amplitude of the power spectrum and making it anisotropic \citep{bharadwaj2004,barkana2005,lidz2007,mao2008,mao2012,majumdar2012}. As was shown by \cite{mao2012}, redshift space distortions can change the spherically-averaged 21-cm power spectrum by up to a factor of around 5 at large spatial scales, meaning that fitting models to data without taking this effect into account could result in significant systematic errors. It is thus important to understand quantitatively how strong these effects are at different scales and at different stages of reionization.

Since redshift space distortions only affect the signal along the line-of-sight, they introduce anisotropies in the otherwise isotropic signal. If the signal-to-noise is sufficient, these anisotropies can be used to remove the complicated astrophysical contribution to the power spectrum and extract pure cosmological information from 21-cm observations \citep{barkana2005,mao2012,shapiro2012}. In situations where the noise level is too high for this to be feasible, it may still be possible to extract the astrophysical information contained in the anisotropies, which is interesting in its own right. 

%Using the quasi-linear approximation of \cite{mao2012}, the 21-cm power spectrum at a fixed $k$ value can be written as a fourth-degree polynomial in $\mu \equiv k_{\parallel}/|\mathbf{k}|$, with the second and fourth moments being determined by the cross-power between neutral and total density and the total density auto power spectrum respectively. We suggest using the sum of the $\mu^2$ and $\mu^4$ terms as a function of redshift as an additional observable of the history of reionization. Since the matter auto power spectrum evolves slowly and predictably, the evolution of this quantity is determined primarily by the (anti-)correlation between neutral and total density. 

The outline of the paper is as follows. In Sec.\ \ref{sec:theory} we briefly summarise the theory behind redshift space distortions and the effects of gas peculiar velocities on the 21-cm power spectrum. In Sec.\ \ref{sec:simulations} we describe our simulations of the 21-cm signal in real- and redshift-space, the instrument noise and the foregrounds. In Sec.\ \ref{sec:results} we show the results from the simulations, demonstrating both the effects of redshift space distortions on the actual 21-cm signal at different scales and global ionization fractions, and the extent to which these effects will be visible in LOFAR observations. We start with a simplified scenario including only instrument noise, and then simulate a more realistic scenario taking into account a larger number of complicating effects such as foreground subtraction. We also discuss how the evolution of the power spectrum anisotropy can be used to constrain the reionization history. Finally, in Sec.\ \ref{sec:summary} we summarise and discuss our results.

For the simulations, we have assumed a flat $\Lambda$CDM model with $(\Omega_{\mathrm{m}}, \Omega_{\mathrm{b}}, h, n, \sigma_8) = (0.27, 0.044, 0.7, 0.96, 0.8)$, consistent with the 9 year WMAP results \citep{hinshaw2012}.

\section{Theory}
\label{sec:theory}
In this section, we go through some of the basic theory of the 21-cm signal that LOFAR will observe. We describe the concept of redshift space, and show how the 21-cm signal differs between real space and redshift space. We also discuss the 21-cm power spectrum, and show how this will be affected by redshift space distortions.

\subsection{21-cm radiation}
Instruments such as LOFAR will attempt to observe the 21-cm emission from neutral hydrogen during the epoch of reionization. 21-cm photons are emitted when the electrons of hydrogen atoms undergo a spin-flip, and the intensity of the radiation depends on the density of neutral hydrogen atoms and the ratio between the two spin populations. This ratio is expressed through the spin temperature $T_s$:
\begin{equation}
	\frac{n_1}{n_0} \equiv \frac{g_1}{g_0} e^{-T_{\star}/T_s},
	\label{eq:spintemp}
\end{equation}
where $n_0$ and $n_1$ are the number densities of atoms in the low and high energy spin states, $g_0=1$ and $g_1=3$ are the statistical weights and $T_{\star} = 0.068$ K is the temperature corresponding to the rest-frame frequency of the transition.

The 21-cm radiation will be observed against the background provided by the CMB. Depending on the temperature of the CMB and the spin temperature, the 21-cm line can be visible either in emission or in absorption. The actual observable quantity is the \emph{differential brightness temperature}, $\delta T_b$. In its most general form, $\delta T_b$ depends on the ratio between the spin temperature and the CMB temperature as well as the velocity gradient along the line-of-sight. Here, we will make two simplifying assumptions: that the line is optically thin (which simplifies the dependency on the velocity gradient, since we can ignore radiative transfer effects), and that $T_s \gg T_{\mathrm{CMB}}$. The first of these assumptions is likely to be true in all but the very earliest stages of reionization, and smallest spatial scales \citep{mao2012}. The assumption of high $T_S$ is somewhat more uncertain in the early stages of reionization (e.g.\ \citealt{mesinger2013}). While full modelling of the $T_S$ fluctuations is beyond the scope of this work, we discuss in Sec.\ \ref{sec:summary} how our results might change at the highest redshifts if this assumption is not true.

With these approximations, we may write the brightness temperature at a position $\mathbf{r}$ and redshift $z$ simply as:
\begin{equation}
	\delta T_b(\mathbf{r}, z) = \widehat{\delta T_b}(z)[1 + \deltahi(\mathbf{r})],
	\label{eq:brightnesstemp_short}
\end{equation}
where $\deltahi(\mathbf{r})$ is the fractional over-density of neutral hydrogen and $\widehat{\delta T_b}(z)$ is the mean brightness temperature at redshift $z$. For an in-depth review of 21-cm physics, see e.g.\ \cite{furlanetto2006b} or \cite{pritchard2012}.

% In its most general form, $\delta T_b$ can be written as:
% \begin{align}
% 	\delta T_b (\mathbf{r}, \nu) &= \frac{3 c^3 A_{10} T_{\star} n_{HI}(\mathbf{r}) a_r}{32 \pi \nu_0^3 H(a_r) |1 + (aH)^{-1} \frac{\ud v_{\parallel}}{\ud r_{\parallel}(\mathbf{r}}|} \nonumber \\
% 	& \times \left[1 - \frac{T_{\mathrm{CMB}(a_r)}}{T_s(\mathbf{r})(1-v_{\parallel}(\mathbf{r})/c} \right]
% 	\label{eq:brighnesstemp_long}
% \end{align}

\subsection{Redshift space distortions}
One interesting aspect of 21-cm observations is that they are effectively three-dimensional. Since the observable is a single emission line, observations tuned to a specific frequency will observe only radiation originating from a specific cosmological redshift (assuming the redshift comes from the expansion of the Universe alone). Mapping the redshift to a position along the line-of-sight makes it possible to reconstruct the \hi distribution in 3D, either to make tomographic images (if the signal-to-noise is high), or to measure statistics such as the power spectrum.

However, the mapping from redshift to line-of-sight position is made imperfect by a number of factors---most importantly the fact that the redshift of an emitter is not only caused by the expansion of the Universe, but also by the emitter's peculiar velocity. We will use the term \emph{redshift space} to denote the space that would be reconstructed by an observer assuming that redshifts are caused purely by Hubble expansion. Without a way of determining peculiar velocities independently, the redshift space is the only space that is observable. A redshift $z$ is translated to a comoving redshift space position $\mathbf{s}$ through the following mapping:
\begin{equation}
	\mathbf{s}(z) = \int_0^{z} \frac{c}{H(z')} \ud z'.
	\label{eq:cdist}
\end{equation}

If the redshift $z$ is caused only by the Hubble expansion, then the redshift space position $\mathbf{s}$ of some emitter will be the same as its comoving real space position $\mathbf{r}$. However, if there is also a peculiar velocity $v_{\parallel}$ along the line-of-sight, then an emitter at position $\mathbf{r}$ in real space will be shifted to a position $\mathbf{s}$ in redshift space:
\begin{equation}
	\mathbf{s} = \mathbf{r} + \frac{1 + z_{\mathrm{obs}}}{H(z_{\mathrm{obs}})} v_{\parallel} (t, \mathbf{r}) \hat{r},
	\label{eq:reddist}
\end{equation}
where $1+z_{\mathrm{obs}} = (1+z_{\mathrm{cos}})(1-v_{\parallel}/c)^{-1}$, $z_{\mathrm{obs}}$ is the observed redshift, and $z_{\mathrm{cos}}$ is the cosmological redshift \citep{mao2012}. In other words, an emitter with a peculiar velocity away from the observer (i.e.\ $v_{\parallel}>0$) will be more redshifted than one with no velocity, and will thus appear to be farther away than is really the case, and vice versa.

\subsection{The 21-cm power spectrum in redshift space}
\begin{figure}
	\begin{center}
		\includegraphics[width=7.8cm]{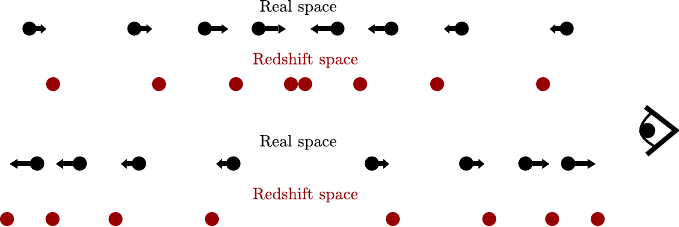}
	\end{center}
	\caption{Illustration of the Kaiser effect, showing how a real-space over-density becomes exaggerated in redshift space. Emitters on the far side of the high-density region will tend to move toward the observer and appear blueshifted, while emitters on the near side will tend to appear redshifted, resulting in dense regions appearing even denser along the line-of-sight. For under-densities, the situation is reversed, as shown in the lower part of the figure.}
	\label{fig:kaiser}
\end{figure}

The motions of gas parcels are not completely random: on average, matter tends to flow towards high-density regions and away from low-density voids. This means that emitters on the far side of a high-density region will tend to appear blue-shifted, and thus closer than they really are, while emitters on the near side will appear farther away. The net effect is that, on the large scales that we are interested in here, dense regions will appear compressed along the line-of-sight, while low-density regions will look emptier than they really are. This effect---illustrated schematically in Fig.\ \ref{fig:kaiser}---is called the \emph{Kaiser effect} \citep{kaiser1987}. It is most familiar from galaxy surveys, but has recently been observed also for intergalactic gas \citep{rakic2012}. While there are other effects that distort positions in redshift space\footnote{For instance, using erroneous values for the cosmological parameters will introduce a $\mu^6$ dependence to the power spectrum (the Alcock-Paczynski effect; \citealt{alcock79,nusser2005}). For this study, we assume that the relevant cosmological parameters are known.}, from now on, for simplicity, we will use the term \emph{redshift space distortions} to mean the shifts in apparent position caused by gas peculiar velocities. With the assumptions we are making here (optically thin line, $T_S \gg T_{\mathrm{CMB}}$), 21-cm redshift space distortions are completely analogous to redshift space distortions in galaxy surveys.

One of the most interesting statistical quantities for LOFAR---and the one we are concerned with in this paper---is the power spectrum of the 21-cm differential brightness temperature fluctuations, or \emph{21-cm power spectrum} for short. The power spectrum $P_{21}(\mathbf{k})$ is defined as:
\begin{equation}
	\langle \widetilde{\delta T_b^*}(\mathbf{k}) \widetilde{\delta T_b}(\mathbf{k'}) \rangle \equiv (2 \upi)^3 P_{21}(\mathbf{k}) \delta_D^3(\mathbf{k}-\mathbf{k'}),
	\label{eq:ps_def}
\end{equation}
where $\widetilde{\delta T_b}$ is the Fourier transform of $\delta T_b$ and $\delta_D^3$ is the three-dimensional Dirac delta function. It is often useful to work with the dimensionless power spectrum, defined as
\begin{equation}
	\Delta^2_{21}(k) = \frac{k^3}{2 \upi^2} P_{21}(k),
	\label{}
\end{equation}
which for the case of the 21-cm power spectrum is in fact not dimensionless, but has the units of mK$^2$.

In real space, the cosmological 21-cm signal is isotropic, meaning that $P_{21}(k)$ is expected to have the same value for all $\mathbf{k}$ with a fixed magnitude $k$. However, in redshift space, this is no longer true---the 21-cm power spectrum will now depend on the direction of $\mathbf{k}$. It is convenient to work with the parameter $\mu$, which is the cosine of the angle between the direction in $k$ space and the line-of-sight, or $\mu \equiv k_{\parallel}/|\mathbf{k}|$. 

The 21-cm power spectrum in redshift space in a spherical shell at a fixed value of $k$ can be written as a fourth-order polynomial in $\mu$. \cite{barkana2005} showed that in the limit of linear fluctuations in density, velocity and ionized fraction, the moments of this polynomial are power spectra of various underlying fields, which are themselves independent of $\mu$. \cite{mao2012} extended this to non-linear ionization fluctuations, but linear density and velocity fluctuations. In this so-called quasi-linear approximation, the power spectrum can be written as:
\begin{equation}
	P_{21}(k, \mu) = P_{\mu^0}(k) + P_{\mu^2}(k) \mu^2 + P_{\mu^4}(k) \mu^4,
	\label{eq:pk_mu}
\end{equation}
where the moments are:
\begin{align}
	&P_{\mu^0} = \widehat{\delta T_b}^2 P_{\deltahi,\deltahi} (k)  \label{eq:mu0_term} \\
	&P_{\mu^2} = 2 \widehat{\delta T_b}^2 P_{\deltahi,\deltah} (k) \label{eq:mu2_term} \\
	&P_{\mu^4} = \widehat{\delta T_b}^2 P_{\deltah,\deltah} (k) \label{eq:mu4_term}
\end{align}
Here, $\widehat{\delta T_b}$ is the mean brightness temperature, and $\delta_x \equiv x/\bar{x}-1$ denotes the fractional over-density of the quantity $x$. Note that the zero-th moment is the power spectrum that would be observed if there were no redshift space distortions in the signal, i.e.\ it is the 21-cm power spectrum in real space. $\deltah$ is the baryonic matter over-density. We assume that this is equivalent to the total matter over-density, which is reasonable on all but the smallest scales, where baryons no longer trace the dark matter. The fourth moment is thus the matter power spectrum familiar from cosmology.

The quasi-linear approximation ignores higher-order terms that become important at small spatial scales and late in the ionization history \citep{shapiro2012}. For the scales we are interested in here, we have found that it provides an adequate approximation for the early stages of reionization ($\langle x \rangle_m \lesssim 0.3$), while comparisons to the quasi-linear approximation at later stages should be interpreted more cautiously.

%%%%%%%%%%%%%%%%%%%%%%%%%%%%%%%%%%%%%%%%%%%%%%%%

\section{Simulations}
\label{sec:simulations}

To simulate the 21-cm signal in redshift space, we used a three-step process. First, an $N$-body simulation was performed to obtain time-evolving density and velocity fields. Then, the reionization of the IGM was simulated through a ray-tracing simulation to get the 21-cm brightness temperature. Lastly, we combined the velocity information from the $N$-body simulations with the brightness temperature to calculate the 21-cm signal in redshift space. We also simulated the instrumental noise for LOFAR observations. To study realistic observations, we simulated galactic and extra-galactic foregrounds. The foregrounds are only applied in the later part of the paper, in Sec.\ \ref{sec:foregrounds}. Below, we describe each of these steps in more detail.

\subsection{Simulations of the 21-cm signal}

\subsubsection{$N$-body and radiative transfer simulations}

The $N$-body simulations were done with \cthree \citep{iliev2008,harnoisderaps2012}, which is built on the \textsc{PMFAST} code \citep{merz2005}. \cthree is a massively parallel hybrid (\textsc{MPI} + \textsc{OpenMP}) particle-particle-particle-mesh code. Forces are calculated on a particle-particle basis at short distances and on a grid for long distances. For this simulation, 5488$^3$ particles with a mass of $5\times 10^7 \Msol$ were used, with a grid size of 10976$^3$. The simulation volume was (607 cMpc)$^3$ (comoving Mpc). For each output from the $N$-body simulations, haloes were identified using a spherical over-density method, resolving haloes down to $\sim 10^9 \Msol$. In addition to this, haloes down to $10^8 \Msol$ were added using a sub-grid recipe calibrated to higher-resolution simulations on a smaller volume.

Each of the outputs from \cthree was then post-processed using the radiative transfer code \ctwo \citep{mellemac2ray} on a grid with $504^3$ cells (i.e.\ a cell size of $1.2$ cMpc) to get the evolution of the ionized fraction. \ctwo uses short-characteristics ray-tracing to simulate the ionization, given some source model. Here, the production of ionizing photons, $\dot{N}_{\gamma}$, for a halo of mass $M_{\mathrm{h}}$ was assumed to be:
\begin{equation}
	\dot{N}_{\gamma} = g_{\gamma} \frac{M_{\mathrm{h}} \Omega_\mathrm{b}}{(10\;\mathrm{Myr}) \Omega_\mathrm{m} m_{\mathrm{p}}},
	\label{eq:ionizing_flux}
\end{equation}
where $m_{\mathrm{p}}$ is the proton mass and $g_{\gamma}$ is a source efficiency coefficient, effectively incorporating the star formation efficiency, the initial mass function and the escape fraction. For this run $g_{\gamma}$ was taken to be:
\begin{align}
	g_{\gamma} = 
	\begin{cases} 
		1.7 & \text{for } M_{\mathrm{h}} \geq 10^9 \Msol \\
		7.1 & \text{for } M_{\mathrm{h}} < 10^9 \Msol.
\end{cases}
\end{align}
These values give a reionization history that is consistent with existing observational constraints \citep{iliev2012}. Sources with $M_{\mathrm{h}} < 10^9 \Msol$ were turned off when the local ionized fraction exceeded 10 per cent, motivated by the fact that these sources lack the gravitational well to keep accreting material in an ionized environment \citep{iliev2007}. The motivation for the higher efficiency of these sources is that they are expected to be more metal-poor than high-mass sources, implying a more top-heavy initial mass function, and a more dust-free environment. The details of this particular simulation will be further explained in Iliev et al.\ (in prep.).

The resulting reionization history reaches $\xm = 0.1$ (global mass-averaged ionized fraction) around $z=9.7$ and $\xm = 0.9$ around $z=6.7$. For the remainder of the paper, we will generally refer to the simulation outputs in terms of $\xm$ rather than redshift, since this makes the evolution of the various physical quantities discussed here slightly less model-dependent.

\subsubsection{Simulating redshift space distortions}
Since our simulations take place in real space, we need some method to transform our data to redshift space. \cite{mao2012} describe a number of ways to calculate the redshift space signal from a real space simulation volume with brightness temperature and velocity information. Here, we use a slightly different method which splits each cell along the line-of-sight into $n$ sub-cells, each with a brightness temperature $\delta T (\mathbf{r})/n$. We then interpolate the velocity and density fields onto the sub-cells, move them around according to Eq.\ \eqref{eq:reddist} and re-grid to the original resolution. This scheme is valid only in the optically thin and high $T_s$ case, when Eq.\ \eqref{eq:brightnesstemp_short} holds and it is possible to treat each parcel of gas as an independent emitter of 21-cm radiation.

\begin{figure*}
	\begin{center}
		\includegraphics[width=18cm]{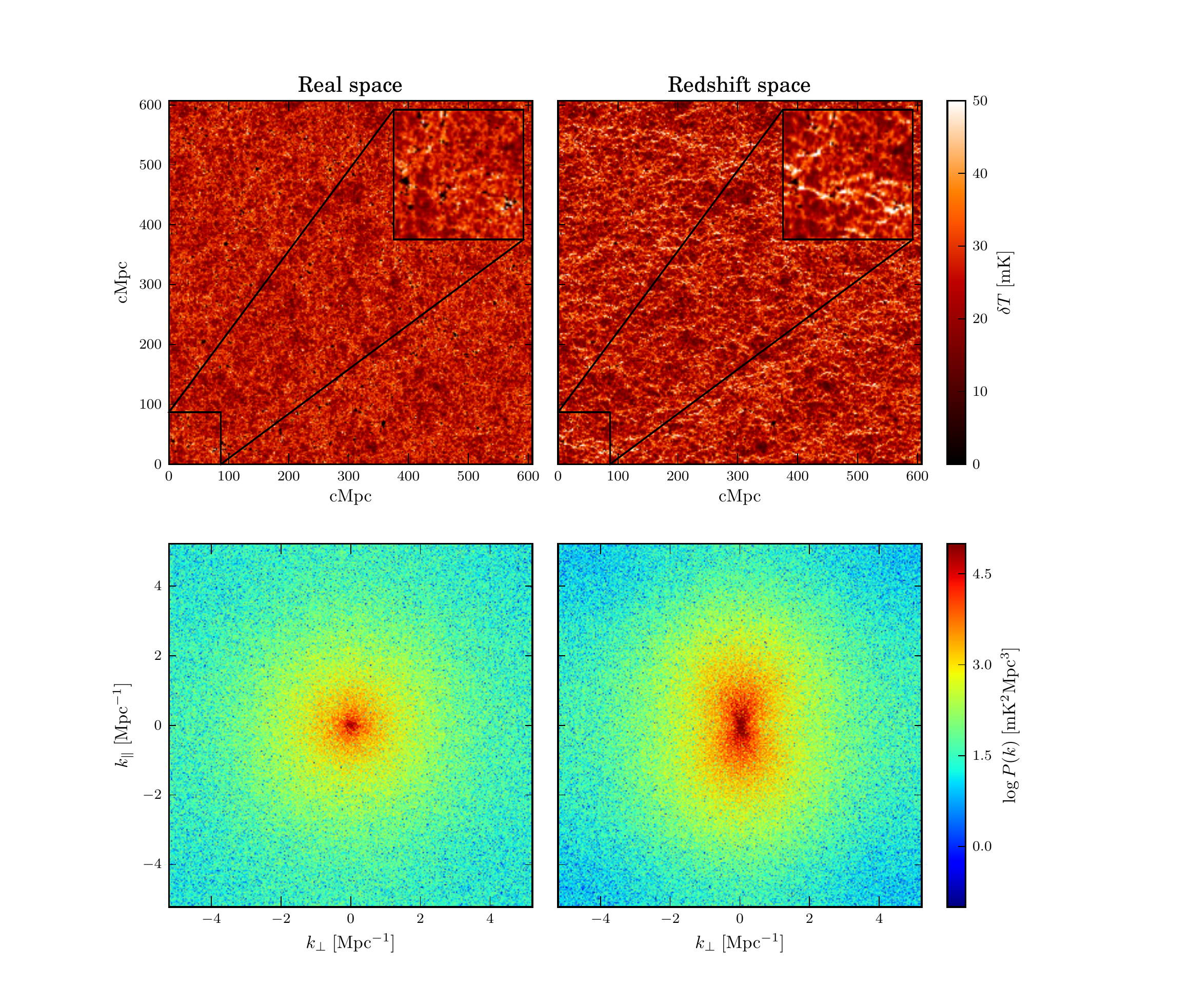}
	\end{center}
	\caption{Visual illustration of the anisotropy and increased contrast induced by redshift space distortions. The top left panel shows a slice through the simulated brightness temperature cube in real space, at $z=9.5$, where $\xm \sim 0.1$. The top right panel shows the same slice, but in redshift space, with the line-of-sight along the $y$-axis. Both panels are $607$ cMpc across, with the bottom-left corners zoomed in to better visualise the increased contrast along the line-of-sight. The bottom panels show slices of the 3D power spectra of the data cubes. Here, the anisotropy in the redshift space signal is clearly visible.}
	\label{fig:pv_effects}
\end{figure*}

This method is similar to the MM-RRM scheme described in \cite{mao2012}, but is simpler to implement and, arguably, more intuitive. We have verified that our results are virtually identical to the MM-RRM scheme for $\gtrsim 20$ sub-cells. For the remainder of the paper we use $50$ sub-cells. As shown in \cite{mao2012}, this method gives valid results at least up to one fourth of the Nyquist wave number, $k_N$. For our simulation volume and resolution, this corresponds to $k \lesssim k_N/4 = \upi/4 \times 504 /(607 \; \mathrm{Mpc}) = 0.65 \; \mathrm{Mpc}^{-1}$. Fig.\ \ref{fig:pv_effects} shows an example slice from our simulations with and without peculiar velocity distortions applied. One can clearly see the anisotropies in redshift space. Regions with a high density in real space will appear to have an even higher density in redshift space, and will look compressed along the line-of-sight. 

\subsection{Simulations of instrumental effects and foregrounds}

\subsubsection{Noise simulations}
To simulate the detector noise contribution to the power spectrum, we use the expression for the RMS noise fluctuation per visibility of an antenna pair, $\Delta V$, found for instance in \cite{mcquinn2006}:
\begin{equation}
	\Delta V = \frac{\sqrt{2}k_B T_{\mathrm{sys}}}{\epsilon A_{\mathrm{eff}} \sqrt{\Delta \nu t}},
	\label{eq:visnoise}
\end{equation}
where $T_{\mathrm{sys}}$ is the system temperature, $A_{\mathrm{eff}}$ is the effective area of the detectors, $\epsilon$ is the detector efficiency, $\Delta \nu$ is the frequency channel width and $t$ is the observing time.

We then make a $u,v$-coverage grid based on the positions of the LOFAR core stations \citep{yatawatta2013} and use Eq.\ \eqref{eq:visnoise} to generate a large number of visibility noise realisations. Each of these realisations is then Fourier transformed to image space, where we apply the same power spectrum calculations as for the signal. Finally, we calculate the standard deviation of the noise power spectra in each $k$ bin to get the noise uncertainty. This procedure is the same as the one described in more detail in \cite{datta2012a}.

\begin{figure}
	\begin{center}
		\includegraphics{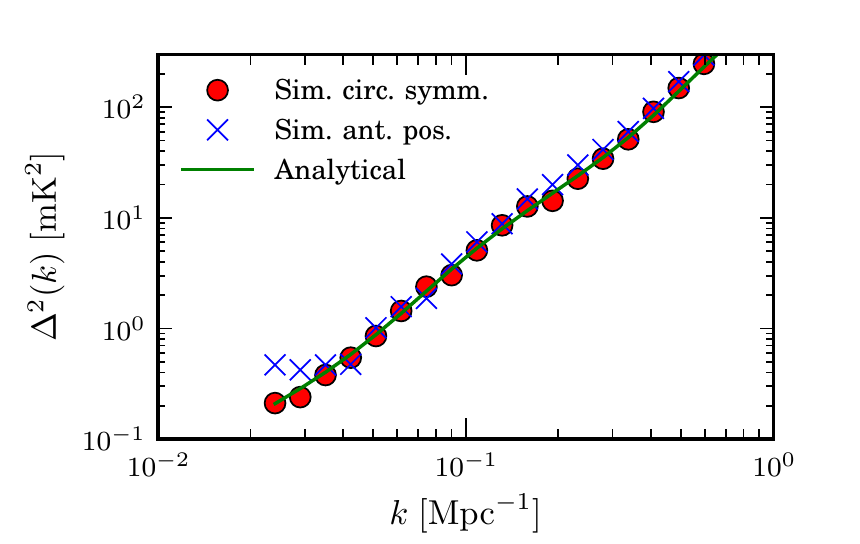}
	\end{center}
	\caption{Simulated and analytically calculated noise power spectrum error, for 1000 hours integration time. The simulations were carried out for the realistic LOFAR core baseline distribution (blue crosses) and for a circularly symmetric $u,v$-distribution (red circles). The error was calculated from 100 noise realisations with $k$ bins of width $\Delta k = 0.19 k$. We also compare to an analytical calculation (green curve). See the text for details, and Tab.\ \ref{tab:noise_parameters} for a list of the parameters used.}
	\label{fig:noise_comparison}
\end{figure}

In Fig.\ \ref{fig:noise_comparison} we show some examples of simulated noise power spectrum errors for 1000 hours integration time. The parameters used for these calculations are listed in Tab.\ \ref{tab:noise_parameters}; see \cite{labropoulos2009} for details. The crosses and circles show the results of Monte Carlo simulations of the noise; each point shows the noise error, calculated as the standard deviation of 100 noise realisations. We show the results for the realistic distribution of the LOFAR core stations (blue crosses) and using a circularly symmetric analytic expression for the $u,v$-distribution (red circles). This analytic expression was chosen to be similar to the realistic baseline distribution and consists of the sum of two Gaussians. Since the realistic baseline distribution was calculated from a 12 hour observation, it is very close to circularly symmetric, and there is very little difference between the two simulations. To demonstrate that the simulated noise errors are reasonable, we also show the results from analytic calculations of the noise error (green curve), using the expressions in \cite{mcquinn2006}. These were calculated for the same instrument parameters as the Monte Carlo simulations, using the double-Gaussian expression for the $u,v$-coverage. 

In general, these same parameters are used throughout the paper unless stated otherwise (with the exception of the observing frequency, which is being varied). We use Monte Carlo simulations of the noise with $u,v$-coverage calculated from the realistic LOFAR core station positions, with a tapering function that cuts off baselines with $|u|>600$. With the tapering, the point-spread function looks similar to a Gaussian with a width of $\approx 3$ arcmin. The tapering does not affect the noise power spectrum at the scales we are interested in, but brings down the per-pixel noise which helps in the foreground removal later on. This results in a noise RMS of $\approx 48$~mK at 190~MHz and $\approx 180$~mK at 115~MHz after 1000 hours of integration.

\begin{table}
	\centering
	\begin{tabular}{ll}
		\hline
		System temperature & $140 \; \mathrm{K}+ 60 \left( \frac{\nu_c}{300 \; \mathrm{MHz}}\right)^{-2.55}  \; \mathrm{K}$  \\
		Effective area  & $526 \left( \frac{\nu_c}{150 \; \mathrm{MHz}} \right)^{-2} \; \mathrm{m^2}$ \\
		Detector efficiency & 1 \\
		Central frequency & 134.8 MHz \\
		Channel width & 0.3 MHz  \\
		Frequency range & 113.2 MHz to 158.2 MHz \\
		Number of stations & 48 \\
		Station beam field-of-view & $5\degree \times 5 \degree$ \\
		$u,v$ coverage & 12 hour observation using LOFAR core \\
		 & stations with source in the zenith\\
		\hline
	\end{tabular}
	\caption{Noise simulation parameters for an observation with central redshift $z_c = 9.5$, as seen in Fig.\ \ref{fig:noise_comparison}.}
	\label{tab:noise_parameters}
\end{table}

\subsubsection{Foreground simulations}
\label{sec:foreground_simulations}
LOFAR observations of the redshifted 21-cm line will be contaminated by foregrounds originating from a number of sources: localised and diffuse Galactic synchrotron emission, Galactic free-free emission and extragalactic sources such as radio galaxies and clusters \citep{jelic2008}. In Sec.\ \ref{sec:foregrounds}, we study the effects of these foregrounds on the observability of redshift space distortions.

The foregrounds were simulated using the models described in \cite{jelic2008,jelic2010}. We do not consider the polarisation of the foregrounds, as recent observations indicate that it should not be a serious contamination for the EoR \citep{bernardi2010}. Furthermore, we assume that bright sources have been accurately removed, for example using directional calibration \citep{kazemi2011}. The foregrounds simulated here can be up to five orders of magnitude larger than the signal we hope to detect but since interferometers such as LOFAR measure only fluctuations, foreground fluctuations dominate by `only' three orders of magnitude (e.g.\ \citealt{bernardi2009}).

The contrast between the smooth spectral structure of the foregrounds and the spectral decoherence of the noise and 21-cm signal lends itself well to a foreground fitting method along the line-of-sight. Though parametric methods such as polynomial fitting have proved popular (e.g. \citealt{santos2005,wang2006,mcquinn2006,bowman2006,jelic2008,gleser2008,liu2009a,liu2009b,petrovic2011,wang2013}), the non-parametric line-of-sight methods so far utilised \citep{harker2009b,chapman2013,chapman2012a,paciga2013} reduce the risk of foreground contamination due to an incomplete model of the foregrounds. 

Here, we choose to remove the foregrounds using a technique called Generalized Morphological Component Analysis, or \textsc{gmca} (\citealt{bobin2007,bobin2008a,bobin2008b,bobin2012}). Initally used for CMB data analysis \citep{bobin2008a}, \textsc{gmca} has been shown to recover simulated EoR power spectra to high accuracy across a range of scales and frequencies \citep{chapman2013}. Due to the extremely low signal-to-noise of this problem, the 21-cm signal is numerically ignored by the method and can be thought of as an insignificant part of the noise. Instead, \textsc{gmca} works by attempting to describe the foregrounds as being made up of different sparse sources by expanding them in a wavelet basis. \textsc{gmca} aims to find a basis set in which the sources to be found are sparsely represented, i.e. a basis set where only a few of the coefficients would be non-zero. With the sources being unlikely to have the same few non-zero coefficients one can then use this sparsity to more easily separate the mixture and remove the foregrounds from the signal.

The full details of the \textsc{gmca} algorithm can be found in \cite{chapman2013} or, outside of the EoR data model, in \cite{bobin2007,bobin2008a,bobin2008b,bobin2012}.

%%%%%%%%%%%%%%%%%%%%%%%%%%%%%%%%%%%%%%%%%%%%%%%%
\section{Results}
\label{sec:results}

\begin{figure*}
	\begin{center}
		\includegraphics{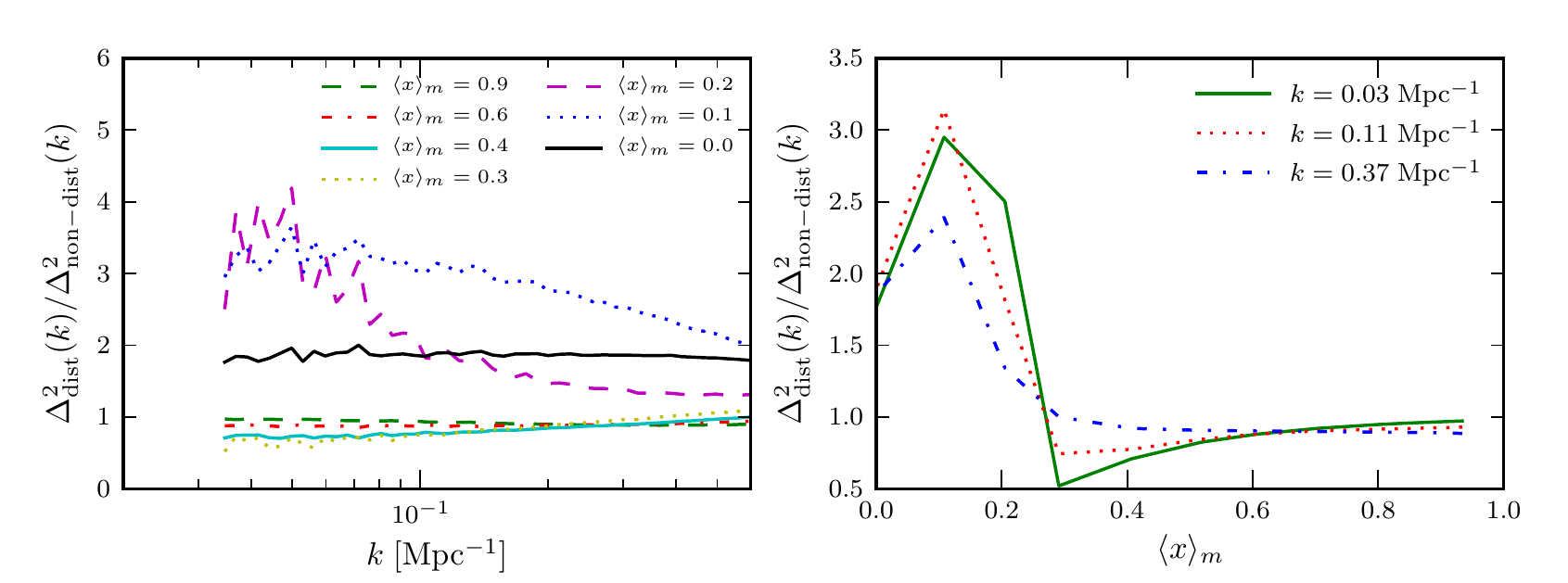}
	\end{center}
	\caption{Ratio between the spherically-averaged power spectrum with the full non-linear redshift space distortions included, and without any redshift space distortions. The left panel shows the ratio for fixed global ionized fractions as a function of $k$. The right panel shows the ratio for fixed $k$ values as a function of ionized fraction.}
	\label{fig:spherical_ratio}
\end{figure*}

In this section, we present the results of our simulations. We begin by quantifying how the 21-cm power spectrum will be distorted in redshift space on various scales and at various stages of reionization. We then investigate to what extent these distortions are visible in LOFAR data, and show how redshift space distortions can be used to constrain the reionization model.

\subsection{Effects of redshift space distortions on the 21-cm power spectrum}

Redshift space distortions modify the 21-cm brightness temperature in two major ways, as is illustrated in Fig.\ \ref{fig:pv_effects}. First, they increase the contrast, which can either amplify or dampen the spherically-averaged power spectrum. Second, they introduce anisotropies into the otherwise isotropic signal.

The effects of redshift space distortions on the spherically-averaged power spectrum were examined in detail by \cite{mao2012}. By averaging Eq. \eqref{eq:pk_mu} over a spherical shell, we get the quasi-linear expectation for the spherically-averaged power spectrum:
\begin{align}
	P_{21}^{\mathrm{qlin}}(k) = &\widehat{\delta T_b}^2 \left[ P_{\deltahi,\deltahi} (k) + \right. \nonumber \\
	& + \left. \frac{2}{3} P_{\deltah,\deltahi} (k) + \frac{1}{5} P_{\deltah,\deltah} (k) \right].
	\label{eq:pk_spherical}
\end{align}
For comparison, the 21-cm power spectrum without redshift space distortions taken into account is given by:
\begin{equation}
	P^{\mathrm{Real\;space}}_{21}(k) = \widehat{\delta T_b}^2  P_{\deltahi,\deltahi} (k).
	\label{eq:pk_spherical_upv}
\end{equation}

\noindent This means that in the earliest stages of reionization, when $\deltahi \approx \deltah$, redshift space distortions amplify the power spectrum by approximately a factor $1+\frac{2}{3} + \frac{1}{5} = 1.87$. \cite{mao2012} showed that the power spectrum is amplified by up to a factor $\sim 5$ in the early stages of reionization, and later on suppressed. In Fig.\ \ref{fig:spherical_ratio} we show the results from our simulations (including the full non-linearities) for the spherically-averaged power spectrum. Note that the ratio stays at approximately 1.87 before reionization starts (black curve), for the large spatial scales plotted here. At smaller scales, the ratio will deviate from this value due to a combination of non-linear effects and the limited resolution of our simulations. For this paper, we are focusing on scales on the order of $k\sim0.1$ Mpc$^{-1}$, where the effects of non-linearities are small.

\begin{figure}
	\begin{center}
		\includegraphics{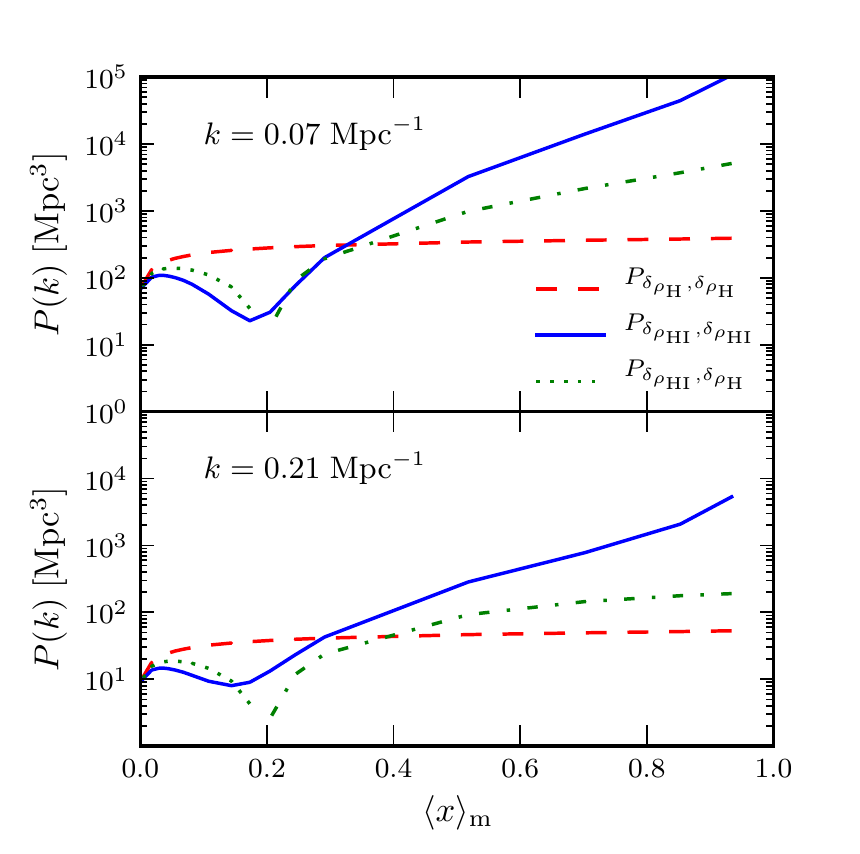}
	\end{center}
	\caption{The three power spectra that affect the $\mu$ dependence of the full brightness temperature power spectrum, as a function of global ionized fraction for two different $k$ modes. Where the H--\hi cross-power spectrum becomes negative, we show $-P_{\deltahi,\deltah}$ as a dot-dashed line. Note that these are the power spectra of the density fluctuations, not the brightness temperature.}
	\label{fig:ps_components}
\end{figure}

To better understand the effects of redshift space distortions on the power spectrum, both in terms of amplification/suppression and in terms of anisotropies, it is illustrative to look at the three power spectra that make up the moments in Eq.\ \eqref{eq:pk_mu}\@. Fig.\ \ref{fig:ps_components} shows the evolution of $ P_{\deltahi,\deltahi} (k)$, $P_{\deltah,\deltahi} (k)$ and $P_{\deltah,\deltah} (k)$ calculated from our simulated data for two fixed $k$ modes as reionization progresses. Each of these power spectra determines one of the $\mu$ terms in the polynomial in Eq.\ \eqref{eq:pk_mu}.

\begin{enumerate}[(i)]
	\item The matter power spectrum, $P_{\deltah,\deltah} (k)$, is the most straightforward of the three, since it depends only on fundamental cosmology and not on the complicated astrophysics of reionization. As over-dense regions accrete matter over time, the matter power spectrum grows monotonically.
	\item The \hi auto-power spectrum, $P_{\deltahi,\deltahi} (k)$, initially follows the matter power spectrum, since almost all hydrogen is neutral early on. As the peaks in the density field become ionized, the \hi auto-power starts to decline. At a global ionized fraction of around 20 per cent, there are almost no large scale \hi fluctuations left, and $P_{\deltahi,\deltahi}$ is negligible compared to the matter power spectrum. After this, the \hi power spectrum turns around and becomes very strong in amplitude. This turn-around occurs when the ionized regions become large enough to provide large-scale fluctuations in \hi. Since $\deltahi$ is defined as the over-density compared to the mean \hi density, the power increases greatly in strength as the mean \hi density decreases.
	\item The H--\hi cross-power spectrum, $P_{\deltah,\deltahi} (k)$, also follows the matter power spectrum initially. Like the \hi auto-power spectrum, it too decreases in strength when the dense peaks become ionized. Since reionization proceeds inside-out in our model (i.e.\ high-density regions tend to ionize before low-density regions), the H and \hi densities will become anti-correlated, and the cross-power spectrum becomes negative (indicated by the dot-dashed lines in Fig.\ \ref{fig:ps_components}).
\end{enumerate}

\begin{figure}
	\begin{center}
		\includegraphics{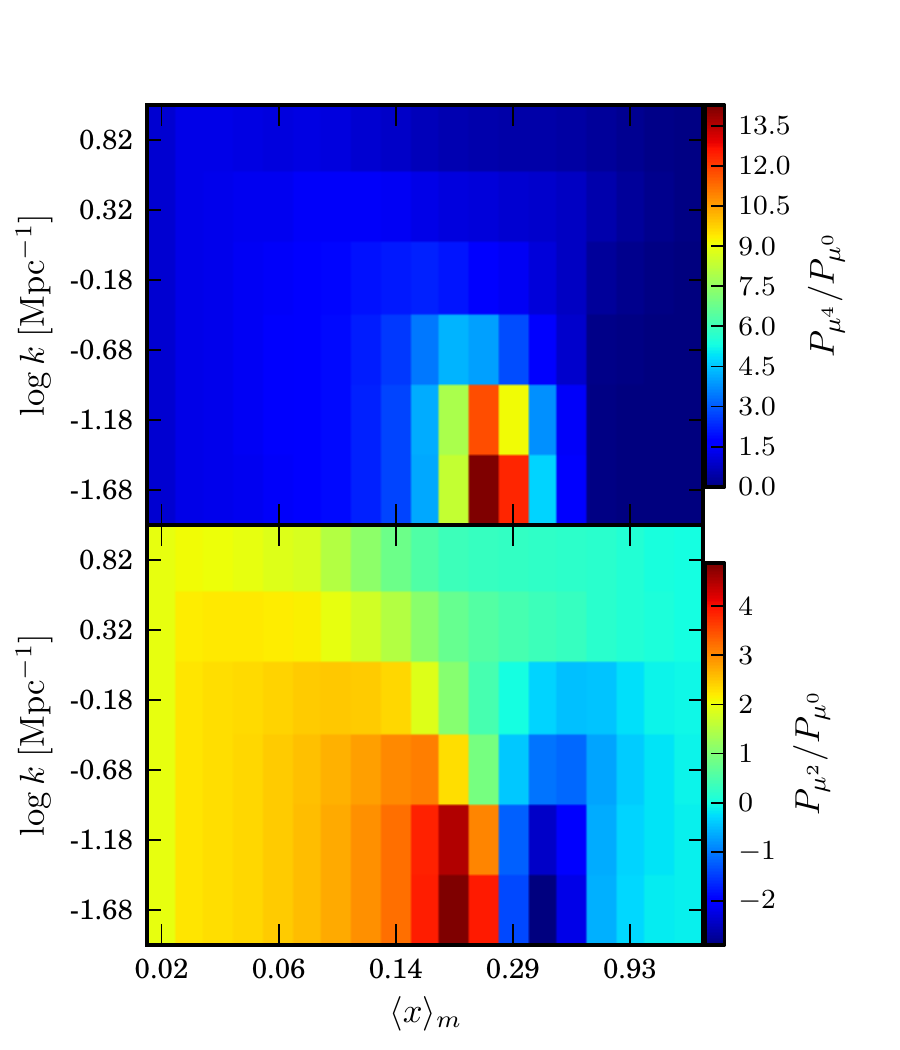}
	\end{center}
	\caption{Evolution of the second and fourth moments of the 21-cm power spectrum, relative to the zeroth moment, for different $k$ modes as reionization progresses. The relative strength of both of the anisotropy terms is highest in the fairly early stages of reionization, and at large spatial scales.}
	\label{fig:terms_ratio}
\end{figure}

\noindent Fig.\ \ref{fig:terms_ratio} illustrates this in a slightly different way. The top panel shows the ratio between the $P_{\mu^4}$ and $P_{\mu^0}$ terms from Eq.\ \eqref{eq:pk_mu} for different $k$ values and global ionized fractions. In the early stages of reionization, the ratio grows slowly as the \hi auto-power spectrum---the $P_{\mu^0}$ term---decreases in strength, while the matter power spectrum continues to grow. At a global ionized fraction of around 20 per cent, the ratio reaches its maximum for large spatial scales. This corresponds to the minimum of the blue line in Fig.\ \ref{fig:ps_components}. After this, the $P_{\mu^0}$ term grows rapidly and the ratio approaches zero.

The bottom panel shows the ratio between the $P_{\mu^2}$ and $P_{\mu^0}$ terms. This ratio also starts off growing since the H--\hi cross-power spectrum decreases more slowly than the \hi auto-power spectrum. However, when the \hi power spectrum turns around and the cross-power spectrum becomes negative, the $P_{\mu^2}/P_{\mu^0}$ ratio rapidly changes sign.

It is clear from Fig.\ \ref{fig:terms_ratio}, that the effects of redshift space distortions are most dramatic at large spatial scales, $k \lesssim 0.2$ Mpc$^{-1}$, and in the rather early stages of reionization, at a global ionized fraction of 10--30 per cent. This is due largely to the suppression of the $P_{\mu^0}$ term that results from the ionization of the highest-density peaks, i.e.\ the dip in the blue curve in Fig.\ \ref{fig:ps_components}.

\begin{figure*}
	\begin{center}
		\includegraphics{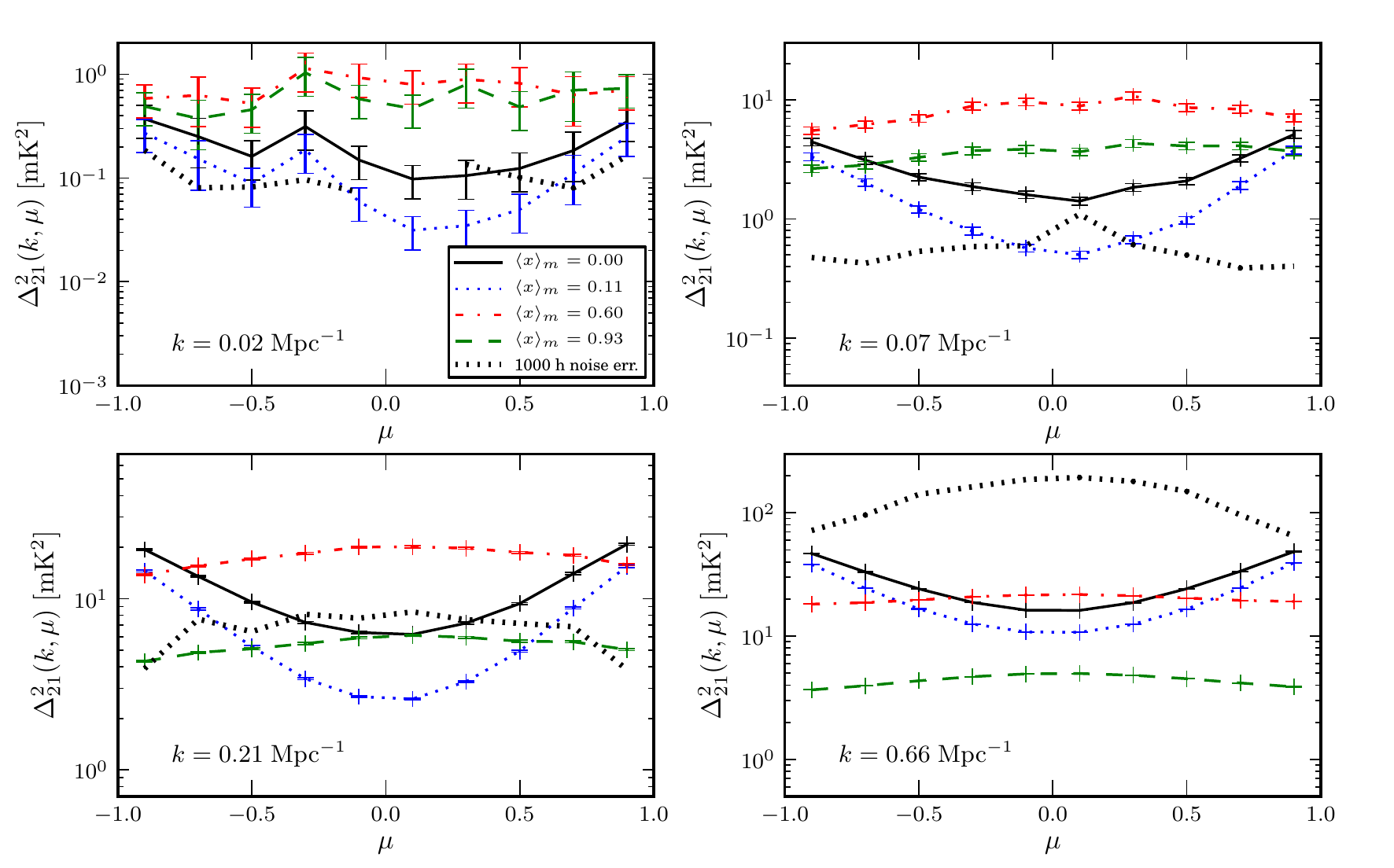}
	\end{center}
	\caption{Power spectrum dependency on $\mu$ at various redshifts for different $k$ modes. Error bars show sample error and the thick dotted lines show the simulated LOFAR noise error at $z=7.2$. All the power spectra were calculated from coeval simulation volumes, i.e.\ evolution effects across the volume are not taken into account.}
	\label{fig:detectability}
\end{figure*}

\subsection{Extraction from simplified mock observations}
Having seen how redshift space distortions alter the 21-cm power spectrum, we now investigate to what extent these effects will be visible in upcoming LOFAR observations. The extraction of the cosmological 21-cm signal from LOFAR measurements will face many hurdles. The signal will be contaminated by factors such as the ionosphere, thermal noise from the instrument and galactic and extragalactic foregrounds. Furthermore, sample errors will limit the interpretation of the largest spatial scales. At small spatial scales and in the later stages of reionization, non-linearities in the density, velocity and ionized fraction fields may spoil the extraction \citep{mao2012,shapiro2012}. Here, we focus first on detector noise and sample error. All the data in this section are calculated from coeval cubes, meaning that we do not take into account any evolution of the signal over the simulation volume---the so-called light-cone effect. In Sec.\ \ref{sec:foregrounds} we study some further complicating factors such as foregrounds. In that section we also include the light-cone effect.

\subsubsection{Signal extraction from noisy data}
Thermal noise from the instrument will add power to the observations on all scales; for LOFAR, the additional power from noise is typically around a factor 10 stronger than the power from the actual signal, for integration times around 500--1000 hours. Two methods have been proposed to extract the signal power spectrum from noisy data \citep{harker2010}. 

The first method relies on knowing the shape of the noise power spectrum, which is a reasonable assumption in a real-world scenario: even if the noise power cannot be calculated theoretically to the required accuracy, it should be possible to measure it empirically. If the signal and noise in a measurement are uncorrelated, then the power spectrum of the noisy signal can be written simply as the sum of the signal power spectrum and noise power spectrum:
\begin{equation}
	P^{\mathrm{signal+noise}}(k) = P^{\mathrm{signal}}(k) + P^{\mathrm{noise}}(k),
	\label{eq:ps_sum}
\end{equation}
and since $P^{\mathrm{noise}}(k)$ is assumed to be known, it can simply be subtracted from the measurements to recover the signal. 

Of course, due to the nature of noise, we can never know its power spectrum exactly, but only the \emph{expectation value}. The expectation value will always have an uncertainty, which we calculate here as the standard deviation of a large number of simulated noise realisations. This noise error, rather than the noise level, is the fundamental limitation to extracting the signal power spectrum from noisy data (although the actual noise level is important in other contexts, such as foreground removal; see Sec.\ \ref{sec:foregrounds}).

The second method does not assume any knowledge of the noise power spectrum. It involves splitting the observing period into two sub-epochs and cross-correlating these. If the signal plus noise in Fourier space for sub-epoch $i$ is $m_i(\mathbf{k}) = s(\mathbf{k}) + n_i(\mathbf{k})$, then the cross-power spectrum between two sub-epochs is:
\begin{align}
	\nonumber
	\langle m_1(\mathbf{k}) m_2^*(\mathbf{k}) \rangle &= \\ \nonumber
	&\langle s(\mathbf{k}) s^*(\mathbf{k}) \rangle + \langle s^*(\mathbf{k}) n_1(\mathbf{k}) \rangle + \\ \nonumber
	&\langle s(\mathbf{k}) n_2^*(\mathbf{k}) \rangle + \langle n_1(\mathbf{k}) n_2^*(\mathbf{k}) \rangle  = \\ 
	 &= \langle s(\mathbf{k}) s^*(\mathbf{k}) \rangle \equiv P^{\mathrm{signal}}(k)
	\label{eq:ps_crosscorr}
\end{align}
since the two noise realisations will be uncorrelated, while the signal is the same. The downside of this method is that foreground subtraction must be carried out separately on the two sub-epochs in order for the cross-terms to vanish \citep{harker2010}. Since each sub-epoch will have lower signal-to-noise than the full data set, this may impact the quality of the foreground subtractions negatively.

To see how well the $\mu$ dependent power spectrum can be extracted from noisy data, we compare in Fig.\ \ref{fig:detectability} our simulated signal to the noise error. Each of the lines show the power spectrum of one of our redshift space distorted brightness temperature cubes. For each of the four panels we have taken the power spectrum at a spherical shell where $|\mathbf{k}| = k$ is fixed and binned it into bins of constant $\mu$. Note that if redshift space distortions were not taken into account, there would be no dependency on $\mu$, and all of these curves would be flat. 

In choosing the width of the bins one is inevitably making a trade-off between an accurate representation of the signal and good noise properties. Here, and for the rest of the paper, we use logarithmic $k$ bins of width $\Delta k = k$. For $\mu$, we use 10 linearly spaced bins. We have found that we need such wide $k$ bins in order to properly reconstruct the signal at reasonable integration times. However, the wide bins will introduce some averaging effects to our signal.

For each $k$ value in Fig.\ \ref{fig:detectability} we show the simulated LOFAR noise power spectrum error for a 1000 hour observation, based on 100 noise realisations (black dotted lines). We also show the sample errors of the signal as error bars. The sample error is calculated as $\Delta_{21}^2(k) \sqrt{2/n}$ where $n$ is the number of Fourier modes that go into the calculation of $\Delta_{21}^2$ at $k$. The $n$ here comes from our simulation volume, which at $z=6.5$ corresponds to approximately $4\times4$ degrees on the sky. The LOFAR beam is similar in size, but will not have full sensitivity over the entire field-of-view. On the other hand, the planned LOFAR observations will eventually comprise several fields, which should bring the sample errors down to levels lower than what we have assumed here.

From Fig.\ \ref{fig:detectability} we see that the noise error completely dominates over the signal on small spatial scales, but goes down for larger spatial scales. The sample error behaves in the opposite way: it is negligible on the smallest spatial scales, but becomes dominant on large scales. At $k$ values somewhere between $0.1$ Mpc$^{-1}$ and $0.2$ Mpc$^{-1}$ (corresponding roughly to angular scales of 10--50 arcmin at $z=8$) there is a sweet spot where both the noise and the sample errors are low enough that the $\mu$ dependency of the power spectrum should be observable with LOFAR, and as Fig.\ \ref{fig:terms_ratio} shows, there are indeed large anisotropies around these $k$ values.

\begin{figure*}
	\begin{center}
		\includegraphics[width=15cm]{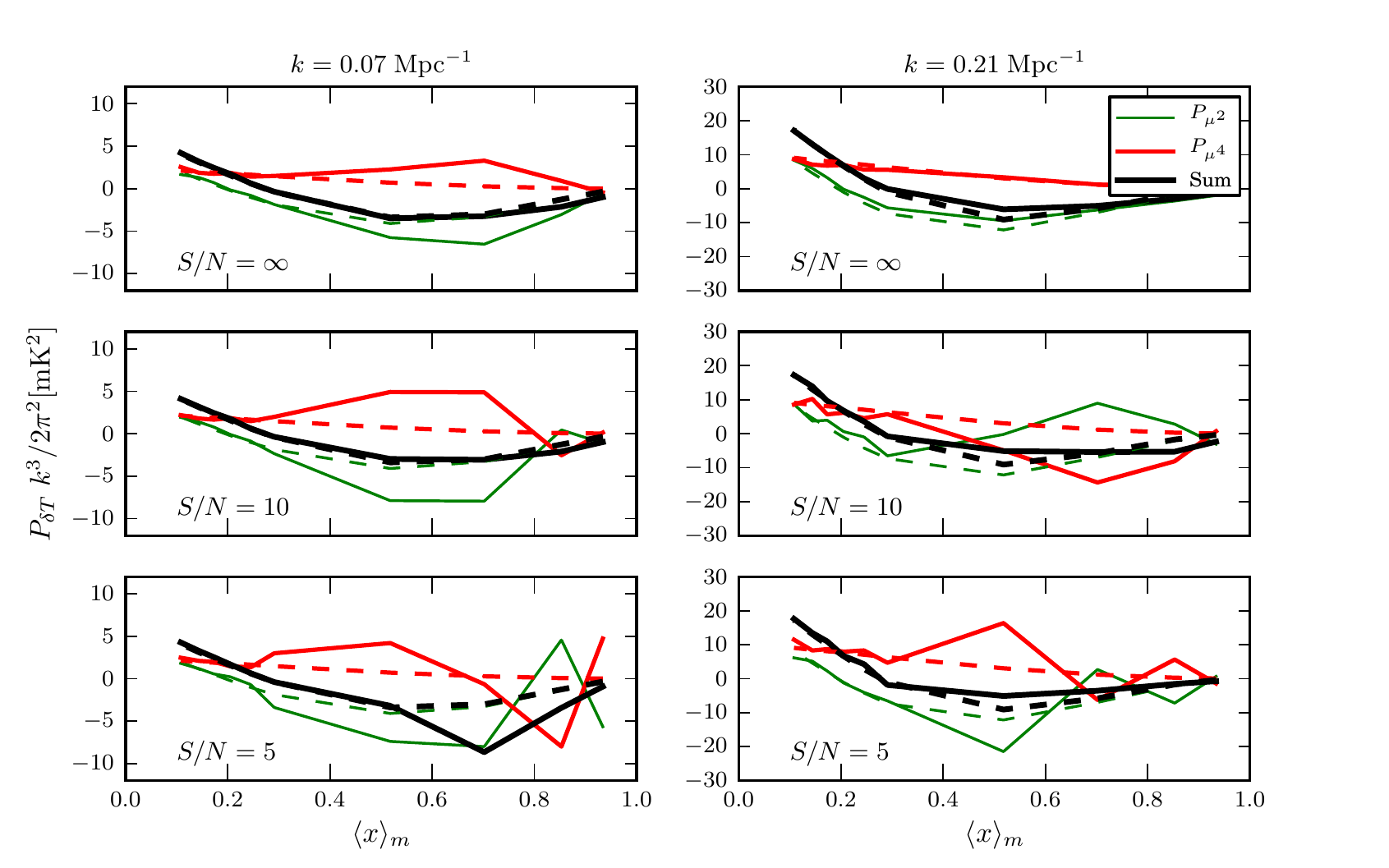}
	\end{center}
	\caption{Leakage of power between the $\mu^2$ and $\mu^4$ terms. The solid curves show extracted terms ($P_{\mu^2}$, $P_{\mu^4}$ and $P_{\mu^2} + P_{\mu^4}$) obtained by fitting polynomials to the signal power spectrum with decreasing signal-to-noise (Gaussian noise added directly to the power spectrum). The top row shows the results for the pure signal, where uncertainties come only from sample errors. The dashed lines show the expectation from the quasi-linear approximation. Notice how the errors on the $\mu^2$ and $\mu^4$ terms are strongly anti-correlated, while the sum of the terms can be extracted much more reliably.}
	\label{fig:term_leakage}
\end{figure*}

\begin{figure*}
	\begin{center}
		\includegraphics{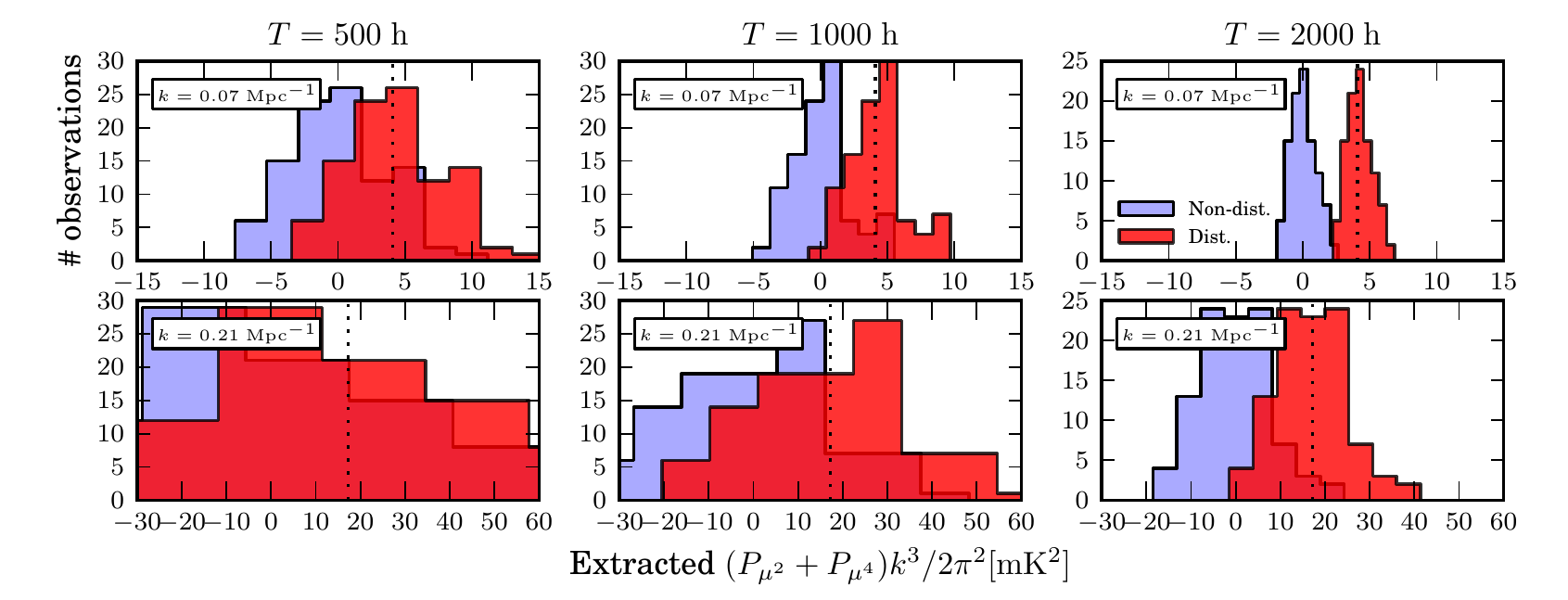}
	\end{center}
	\caption{Extracted anisotropy terms for $k=0.07$ Mpc$^{-1}$ (upper row) and $k=0.21$ Mpc$^{-1}$ (lower row) after 100 mock observations at $z=9.5$, where $\xm = 0.1$. For each mock observation, we extracted the signal by subtracting the expected noise power spectrum from the full power spectrum (see the text for details). The red histograms show the distribution of the extracted anisotropy terms $P_{\mu^2}+P_{\mu^4}$ for LOFAR observations of 500 (left), 1000 (middle) and 2000 hours (right). As a comparison, we show the same thing for the signal with no redshift space distortions included (blue histograms). The dotted lines show the expectation from the quasi-linear approximation. It appears that the anisotropy in the signal can be observed with LOFAR after $\gtrsim 1000$ hours of observations.}
	\label{fig:extracted_terms}
\end{figure*}

To test whether this extraction works, we generated 100 mock observations with different noise realisations but with the same underlying signal. For each such signal+noise cube, we calculated the power spectrum and binned it in $k$ and $\mu$. We then subtracted the expected noise power spectrum---calculated as the mean of many noise realisations---to get the signal power spectrum, according to Eq.\ \eqref{eq:ps_sum}. For each extracted signal power spectrum, we took a specific $k$ bin and fit a fourth-degree polynomial in $\mu$ using a standard least-squares fit with the ansatz that $P_k(\mu) = P_{\mu^0} + P_{\mu^2}\mu^2 +P_{\mu^4}\mu^4$. Ideally, we would expect the terms of this polynomial to correspond to the terms of Eq.\ \eqref{eq:pk_mu}.

In general we find that for all but the lowest noise levels, it is near-impossible to separate $P_{\mu^2}$ from $P_{\mu^4}$, since the two terms tend to leak into each other. This is demonstrated in Fig.\ \ref{fig:term_leakage}, where we have taken the $\mu$-decomposed power spectra at a few different ionization stages and scales, added Gaussian noise to get the specified signal-to-noise ratios, and fitted polynomials as discussed above. We show the individual terms of the fits along with the sum of the second and fourth moments. It is clear that even at these very low noise levels, the errors on the individual terms are very large (compare the signal-to-noise here to the simulated LOFAR noise error of Fig.\ \ref{fig:detectability}, which is of the same order as the signal, i.e.\ about a factor 5 worse than the worst case shown in Fig.\ \ref{fig:term_leakage}). However, the sum of the two terms is much more resilient to the noise, and follows the quasi-linear expectation rather well. The leakage of the two terms into each other is a direct consequence of the fact that $\mu^2$ and $\mu^4$ form a non-orthogonal basis.

Returning to the mock observations, we focus on the extracted sum of the anisotropy terms rather than the individual terms themselves. This is shown in Fig. \ref{fig:extracted_terms}, where the histograms show the sum of the extracted terms for all the mock observations. For comparison, we also show the case where redshift space distortions are not included, in which case we expect $P_{\mu^2}$ and $P_{\mu^4}$ to be zero. At 500 hours, the anisotropy does affect the signal, but the effect is of a similar magnitude as the uncertainty due to the noise. After 1000 hours, there is a fairly strong anisotropy at $k=0.21$ Mpc$^{-1}$, and for $k=0.07$ Mpc$^{-1}$, virtually all mock observations show a positive anisotropy. For 2000 hours, it appears the anisotropy should be clearly visible in the signal for both of the $k$ values we consider here.

%But already for a 1000-hour observation, there is not a single case where we do not see any anisotropy in the reconstructed signal power spectrum, and after 2000 hours, there is very little overlap between the distorted and non-distorted distributions.

\subsection{Redshift space distortions as a probe of reionization}
In the previous section we established that LOFAR should be able to detect redshift space distortion anisotropies in the 21-cm power spectrum after $\gtrsim$ 1000 hours of observations. This fact in itself can be useful as a sanity check for future observations---seeing anisotropy in the signal will greatly increase the credibility of a claimed 21-cm detection. However, it is also interesting to explore whether the anisotropies can be exploited by future observations to obtain additional information, as a complement to the spherically-averaged power spectrum.

It has been suggested that redshift space distortions can be used to probe fundamental cosmological parameters by extracting the matter power spectrum, $P_{\deltah,\deltah} (k)$ \citep{barkana2005,mcquinn2006}. In principle, this is possible by fitting a polynomial in $\mu$ to the power spectrum at a fixed $k$, like we did in Fig.\ \ref{fig:extracted_terms}, and looking only for the $\mu^4$ term. 
In \cite{shapiro2012}, we explored this extraction using the same $N$-body and reionization radiative transfer
simulations to produce our mock 21-cm signal data as used here, but in the limit where sampling errors dominate over noise.
In practice, however, we find that for LOFAR observations, the noise levels are far too high to reliably separate the $\mu^2$ and $\mu^4$ terms.

However, as we saw in Fig. \ref{fig:extracted_terms}, we can extract the sum of the two anisotropy terms, $P_{\mu^2}+P_{\mu^4}$. This sum contains both cosmological and astrophysical information (cf.\ Eqs. \ref{eq:mu2_term} and \ref{eq:mu4_term}), and is not straightforward to interpret. Nevertheless, its evolution with redshift may still tell us something about the history of reionization.

From Fig.\ \ref{fig:ps_components} it is clear that the matter power spectrum---which controls the $\mu^4$ term---evolves rather slowly and predictably with time. This is to be expected, since it does not depend on the complicated astrophysics of reionization, but only on the slow and steady growth of matter over-densities. The H--\hi cross-power spectrum, on the other hand, evolves rapidly with time. We may therefore expect that most of the change in $P_{\mu^2}+P_{\mu^4}$ with redshift will be due to the $\mu^2$ term.

\begin{figure}
	\begin{center}
		\includegraphics[width=\columnwidth]{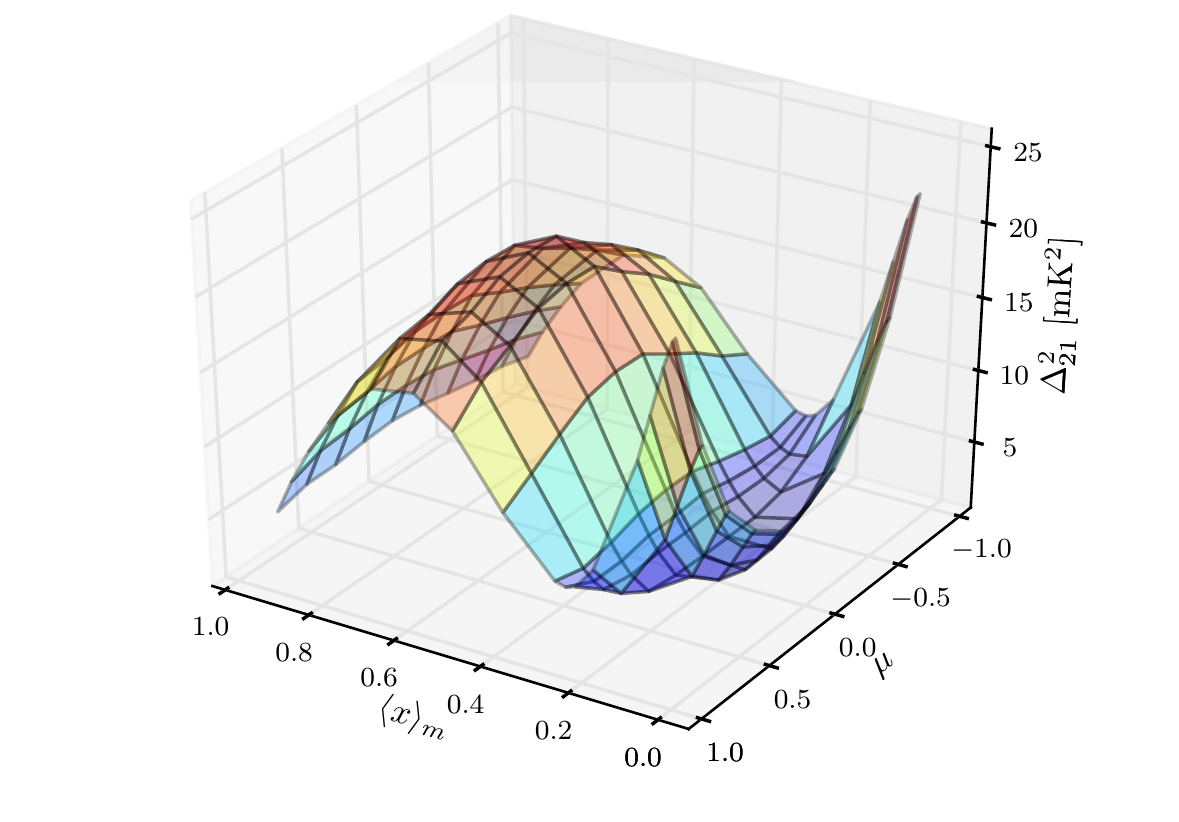}
	\end{center}
	\caption{The evolution of the 21-cm power spectrum as a function of $\mu$ at $k=0.21$ Mpc$^{-1}$ as reionization progresses. Initially, the $\mu^2$ and $\mu^4$ terms both contribute to give a strong positive dependence on $|\mu|$, but as the high-density regions ionize, the $\mu^2$ term becomes negative and the power spectrum flattens, and eventually takes on a negative $|\mu|$ dependence.}
	\label{fig:curvature}
\end{figure}

The evolution of the $\mu$-dependency of the 21-cm power spectrum at $k=0.21$ Mpc$^{-1}$ is shown in Fig. \ref{fig:curvature}. In the early stages of reionization, both $P_{\mu^2}$ and $P_{\mu^4}$ are positive, resulting in a strong positive dependence on $|\mu|$. As the densest areas ionize, $P_{\mu^2}$ drops in strength and eventually becomes negative. At a global ionized fraction of $\xm \approx 0.25$, we have $P_{\mu^2} = -P_{\mu^4}$, and the power spectrum is almost independent of $\mu$. After this, the negative $P_{\mu^2}$ starts to dominate, and the power spectrum changes to a negative dependence on $|\mu|$. The general shape of the surface in Fig.\ \ref{fig:curvature} is the same also for smaller values of $k$.

\begin{figure}
	\begin{center}
		\includegraphics{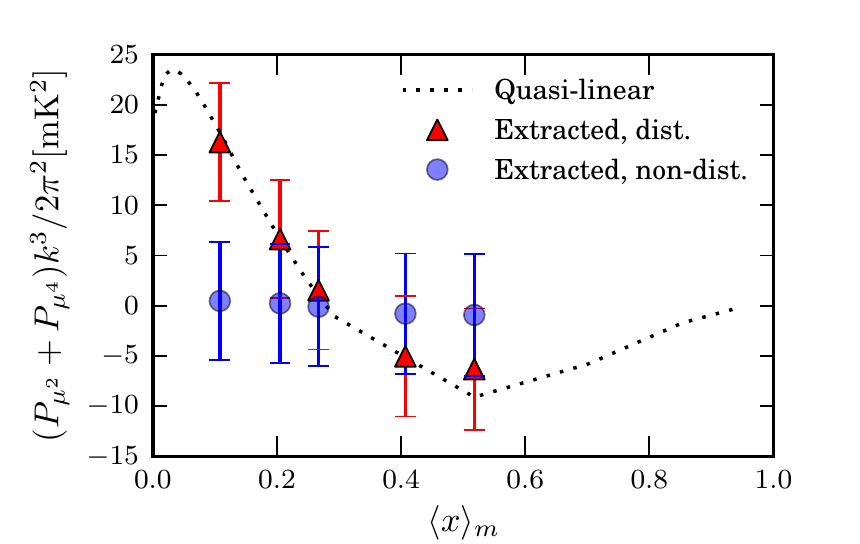}
	\end{center}
	\caption{Reconstructed $P_{\mu^2}+P_{\mu^4}$ at $k=0.21$ Mpc$^{-1}$ as a function of ionized fraction with (red) and without (blue) redshift space distortions included; error bars indicate the standard deviation from 100 noise realisations, calculated for 2000 hours of observation per redshift. The dotted line shows the expectation from the quasi-linear approximation.}
	\label{fig:terms_sum_history}
\end{figure}

In Fig.\ \ref{fig:terms_sum_history}, we show the results from many mock observations like those in Fig.\ \ref{fig:extracted_terms}, for 2000 hours observing time, where we extracted the sum of the anisotropy terms for a number of global ionized fractions. As the dotted line (expectation from the quasi-linear approximation) shows, the sum of the terms is a good indicator of the curvature of the power spectrum, shown in Fig.\ \ref{fig:curvature}. We also see that LOFAR observations should allow us to see the evolution of the curvature with relatively strong certainty.

The exact shape of Fig.\ \ref{fig:curvature} and Fig.\ \ref{fig:terms_sum_history} depends on the reionization model, but some general conclusions can still be drawn. The fact that the sum of the anisotropy terms (the red triangles in Fig.\ \ref{fig:terms_sum_history}) go from positive to negative---and that the curve in Fig.\ \ref{fig:curvature} goes flat and changes curvature---is a direct consequence of reionization progressing inside-out, i.e.\ high-density regions ionizing before low-density regions. Since the matter auto-power spectrum grows monotonically over time, the only way to get negative anisotropy is if the H and \hi densities are sufficiently spatially anti-correlated so that $P_{\mu^2} < - P_{\mu^4}$. The redshift where the anisotropy changes sign will be an indicator of how strongly inside-out reionization is, i.e.\ how strong the anti-correlation between total and neutral density is. 

\begin{figure}
	\begin{center}
		\includegraphics{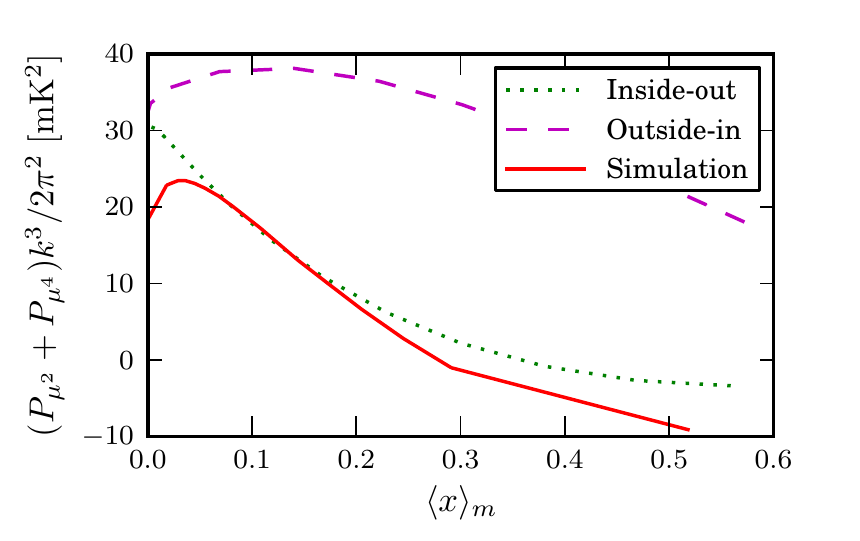}
	\end{center}
	\caption{Evolution of the anisotropy at $k=0.21$ Mpc$^{-1}$ for two toy models representing the two most extreme cases of reionization topology (see the text for details). We also show the results from the \ctwo simulations. While the inside-out model behaves similarly to the simulations, the outside-in model is completely different and never obtains a negative anisotropy.}
	\label{fig:toy_models}
\end{figure}

To further illustrate how the anisotropy evolution depends on the reionization scenario, we show the results from two simple toy models in Fig.\ \ref{fig:toy_models}. Both of these models were constructed from the same, time-evolving, density field as the simulation discussed above. For the first model, labelled ``inside-out'', we assumed that the ionized fraction $x_i$ was:
\begin{align}
	x_i = 
	\begin{cases} 1 & \text{where $\rho > \rho_{\mathrm{th}}$,} \\
	0 &\text{elsewhere,}
\end{cases}
\end{align}
for some threshold density $\rho_{\mathrm{th}}$. The other model, labelled ``outside-in'' represents the other extreme. Here, we put $x_i = 1$ for cells where $\rho < \rho_{\mathrm{th}}$ and 0 everywhere else, similar to what was proposed in \cite{miraldaescude2000}. The threshold density was set at each redshift so that both toy models would get the same mass-averaged ionized fraction as the \ctwo simulations.

As Fig.\ \ref{fig:toy_models} shows, the two models give very different histories for the anisotropy. For the outside-in model, the anisotropy stays at a high level, and only decreases at the later stages because the global $\delta T_b$ decreases. It never becomes negative. The inside-out model, on the other hand, gives an anisotropy evolution that is very similar to the simulations; in fact the anti-correlation appears less extreme. While the inside-out toy model has perfect anti-correlation between matter density and neutral \emph{fraction}, the $\mu^2$ term is determined by the cross-power spectrum of the matter density and the neutral \emph{density}, and the anti-correlation between these two quantities is stronger in the simulated model. Comparing this to the errors bars in Fig.\ \ref{fig:terms_sum_history}, it appears to be well within the capabilities of LOFAR to distinguish between an inside-out and an outside-in model and to exclude at least the most extreme versions of outside-in reionization. 

\subsection{Extraction from more realistic mock observations}
\label{sec:foregrounds}

In the previous sections we have analysed the extraction of anisotropies in a somewhat simplified manner. The data points in Fig.\ \ref{fig:extracted_terms}, for example, all come from output cubes directly from our simulations, with noise that was generated at a single observing frequency. In real observations, a number of factors complicate the extraction of the $\mu$-dependent power spectrum, including:
\begin{enumerate}[(i)]
	\item The light-cone effect. Observations at different redshifts will see the cosmological signal at different evolutionary stages, making the signal at the low-frequency part of an observation different from the high-frequency part \citep{datta2012b}. This affects the average power in a given $k$ bin, but analysis has shown that the extra anisotropy introduced by the light-cone effect is small (Datta et al., in prep).
	\item Frequency dependence of the noise. Since the effective area and system temperature of the telescope depend on the observing frequency, so does the noise level (see Tab.\ \ref{tab:noise_parameters}). 
	\item Resolution effects. The point-spread function of the telescope smooths the signal in the plane of the sky. This can introduce a small $\mu$ dependence in the power spectrum at scales smaller than the resolution. In general, the shape and size of the point-spread function are also frequency dependent.
	\item Angular coordinates. By necessity, observations use angular coordinates on the sky, and frequency along the line-of-sight. To reconstruct the power spectrum, we need to convert the signal to physical coordinates, which will introduce some interpolation effects.
	\item Foregrounds. The signal will be contaminated by several sources of foregrounds. While sophisticated algorithms to remove these exist, the signal will still be degraded somewhat.
\end{enumerate}

\begin{figure*}
	\begin{center}
		\includegraphics[width=17cm]{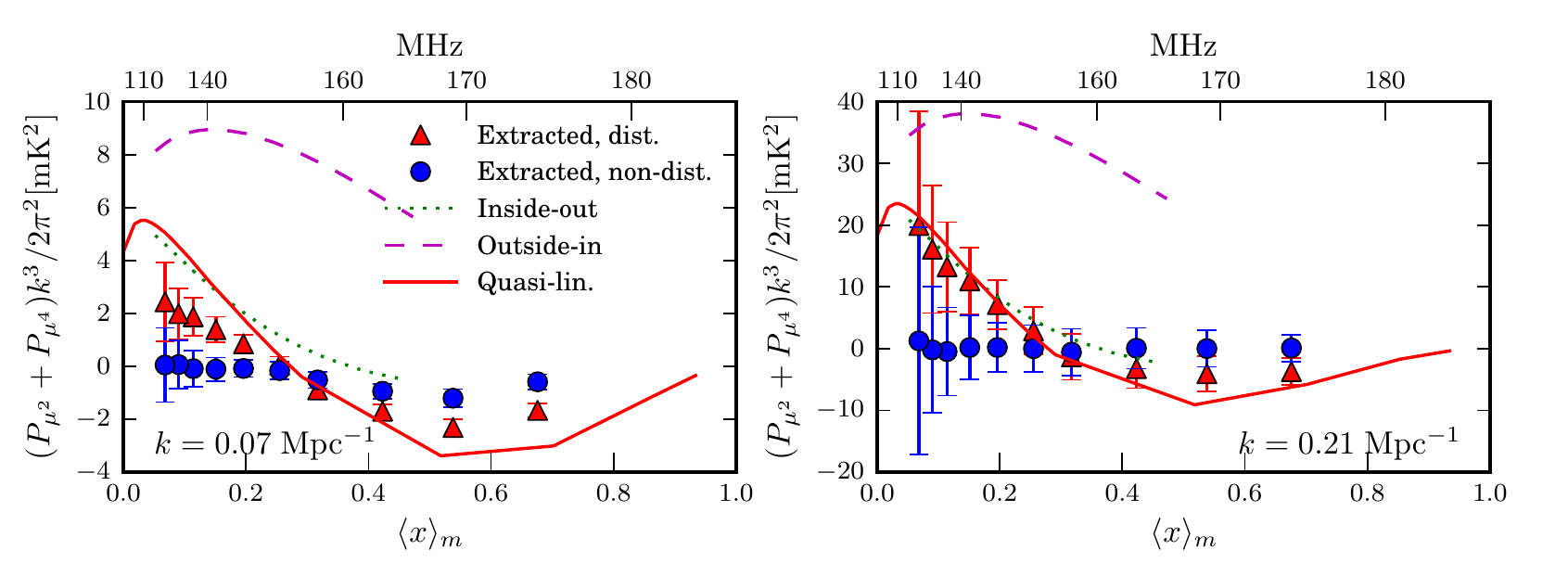}
	\end{center}
	\caption{Extracted terms from light-cone cubes with frequency-dependent noise. Each red point shows the extracted sum of anisotropy terms for slices of 20 MHz depth. The extraction was done by cross-correlating two different noise realisations of 2000 hours integration time each. The error bars show the 1 $\sigma$ spread for 100 different noise realisations. For reference, we also show the same extraction without including redshift space distortions (blue points), as well as the expectations from the quasi-linear model and the two toy models described in the text (dotted lines). The upper $x$-axis shows the observing frequency corresponding to the global ionized fraction in the particular reionization model used in this paper.}
	\label{fig:realistic_extracted}
\end{figure*}

\noindent In this section, we attempt to address these issues\footnote{Further complicating factors, which we do not take into account here, include distortions by the ionosphere and radio frequency interference \citep{offringa2013}. We also do not attempt to model the frequency dependence of the point-spread function.} by generating more realistic mock observations. For this, we created a light-cone cuboid by taking the appropriate slices from our coeval simulation cubes (i.e.\ cubes at a single instant in time) at 0.5 MHz intervals and interpolating between these. See \cite{datta2012b} for more details on the method used. We then smoothed each frequency slice of the light-cone cuboid with a 3 arcmin Gaussian to mimic the LOFAR point-spread function. Finally, we made 100 different noise realisations with the frequency dependence detailed in Tab.\ \ref{tab:noise_parameters}.

To extract the signal power spectrum, we divided the cuboid into slices of 20 MHz depth and for each slice we calculated the cross-power spectrum between two different noise realisations, corresponding to 2000 hours each. This gives the signal auto-power spectrum according to Eq. \eqref{eq:ps_crosscorr}. We then fitted polynomials to the extracted power spectra to get the sum of the anisotropy terms like before. The results of this extraction are shown in Fig.\ \ref{fig:realistic_extracted}, for two different $k$ modes. For reference we show the same extraction when not including redshift space distortions (blue points). We also show the expected anisotropy from the quasi-linear approximation and the two toy models from Fig.\ \ref{fig:toy_models}. There appears to be some bias in the extraction of the anisotropy at $k=0.07$ Mpc$^{-1}$, causing it to deviate from the quasi-linear expectation. This is most likely due to the fact that each extracted power spectrum includes data from a range of frequencies, which introduces some averaging effects, particularly at large scales.

In Fig.\ \ref{fig:realistic_extracted}, the effects of the frequency dependence of the noise are obvious: the extraction is much more uncertain at low frequencies. The details of this uncertainty are highly model-dependent, however. Along the upper $x$-axis we show the observing frequency corresponding to a given global ionized fraction in this particular reionization simulation, but different source models may give significantly later or earlier reionization histories (e.g.\ \citealt{iliev2012}). In general, a late reionization scenario will be easier to observe since that will put the important changes in the signal at frequencies where the noise is lower.

\begin{figure}
	\begin{center}
		\includegraphics{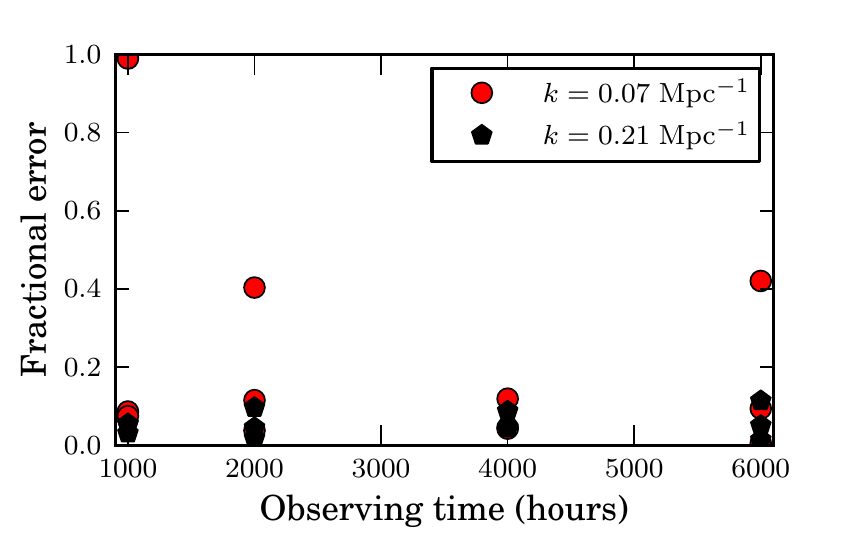}
	\end{center}
	\caption{Fractional contribution to the error in the reconstructed anisotropy terms due to foregrounds at 140 MHz ($z\approx 9$). The errors were calculated by extracting the signal power spectrum using cross-correlation and fitting the $\mu$ polynomial before and after adding and subtracting foregrounds (see the text for details). Each point shows the fractional error for one noise realization.}
	\label{fig:foreground_error}
\end{figure}

\subsubsection{Foregrounds}
To investigate the effects of foregrounds on the anisotropy extraction we added simulated galactic and extragalactic foregrounds to the light-cone cuboid, and used a wavelet method to remove them, as described in Sec.\ \ref{sec:foreground_simulations}. We show the errors due to foregrounds in Fig.\ \ref{fig:foreground_error}. The errors were calculated by carrying out the same anisotropy extraction as in Fig.\ \ref{fig:realistic_extracted} for a number of noise realisations before and after foreground subtraction and removal (only the noise changes between the realizations, the foregrounds were kept the same). The fractional error was then defined as the difference between the extracted sum of terms with and without foregrounds included, divided by the sum of terms without foregrounds included. 

%As Fig.\ \ref{fig:foreground_error} shows, the errors from foregrounds can be significant (around 20 per cent) at the largest spatial scales and short integration times. Increasing the integration time can bring the errors down to less than 10 per cent. For $k=0.21$ Mpc$^{-1}$ the errors are only a few percent, and the effects of foregrounds are small compared to that of the noise. After a few thousand hours of integration, it appears that we hit a point of diminishing returns and the foreground subtraction no longer improves (in fact, from Fig.\ \ref{fig:foreground_error}, it appears to get slightly worse, but since these errors are based on very few samples it is hard to draw any firm conclusions).
As Fig.\ \ref{fig:foreground_error} shows, it seems that foregrounds are manageable at $k=0.21$ Mpc$^{-1}$ even after only $\sim$1000 hours of observation, with no realization adding a bigger error than around 10 per cent. For the largest scale considered here, $k=0.07$ Mpc$^{-1}$, the situation looks a bit worse, with some realizations giving errors up to several tens of percent even at long observing times. This is in line with what we expect: in general foreground subtraction will work best on intermediate scales, since at large scales there will be some leakage of foregrounds into the signal, while at small scales some noise will leak into the signal \citep{chapman2012a,chapman2013}.

While noise is a fundamental limitation to the extraction of the signal, the errors from the foregrounds depend on the method used for foreground subtraction (and possibly also on the input parameters for the method). Several subtraction algorithms exist (e.g.\ \citealt{jelic2008,harker2010,liu2011,chapman2012a,chapman2013}), and it is not obvious that the one we have used here is the optimal method for extracting anisotropies. We will investigate the systematic effects of foregrounds on observing redshift space distortions in a follow-up paper (Chapman et al., in prep). Meanwhile, Fig.\ \ref{fig:foreground_error} is a proof-of-concept that foregrounds can be dealt with efficiently enough to observe the anisotropies, at least for certain scales and frequency ranges.

%%%%%%%%%%%%%%%%%%%%%%%%%%%%%%%%%%%%%%%%%%%%%%%%
\section{Summary and Discussion}
\label{sec:summary}
Observations of the 21-cm emission from the epoch of reionization will inevitably be distorted by the peculiar velocities of the gas in the intergalactic medium. A detailed understanding of how such redshift space distortions affect the 21-cm signal will be crucial for interpreting future observations. We have simulated the effects of redshift space distortions on the 21-cm power spectrum from the epoch of reionization, specifically focused on the anisotropy introduced in the signal. As was already seen in \cite{mao2012}, redshift space distortions strongly affect the 21-cm power spectrum on large scales in the early stages of reionization (around 10--30 per cent global ionization fraction). Here, we have focused specifically on the evolution of the anisotropy. We have shown how, for our reionization model, the power spectrum becomes highly anisotropic in the early stages of reionization , particularly at large spatial scales ($k \sim 0.1$ Mpc$^{-1}$). Initially, the power spectrum has a positive dependence on $|\mu| \equiv |k_{\parallel}|/|\mathbf{k}|$ for the small values of $k$ considered here. As reionization progresses, the increased anti-correlation between the neutral and total matter densities causes the power spectrum to flatten, and eventually depend negatively on $|\mu|$.

We have also studied the observability of the anisotropies with LOFAR. The range of scales around $k \approx 0.07$--$0.2$ Mpc$^{-1}$ (corresponding roughly 10--50 arcminutes on the sky) seems most promising as both the instrument noise and the sample errors are low enough, and foregrounds can be removed with decent accuracy. These scales also happen to correspond to the region in $k$ space with the strongest anisotropies. A $\gtrsim$ 1000 hour observation with LOFAR should reveal anisotropies in the power spectrum, unless the reionization history is significantly different from the scenario in our simulations (for example, a very early reionization would put most of the anisotropies in a frequency range where the noise is higher than we have assumed here). %In fact, these estimations are somewhat conservative, since our values for the sample error come from our simulation volumes which are slightly smaller than the LOFAR beam. Furthermore, LOFAR will observe the EoR signal in several fields simultaneously, which will effectively reduce the sample error by a factor of the square root of the number of fields.

The mere detection of anisotropies in the 21-cm power spectrum would be useful as a check to make sure that the detected signal is indeed the signal from the EoR. As can be seen in Fig.\ \ref{fig:extracted_terms}, it would be highly unlikely to observe for 2000 hours and not detect any anisotropy in the power spectrum. This also shows that when fitting a model to an observed 21-cm power spectrum, it is important to include the effects of redshift space distortions in the model. As Fig.\ \ref{fig:spherical_ratio} indicates, failure to do so may result in systematic errors of several hundred percent. Alternatively, one may use an extraction scheme, such as the one shown in this paper, to remove the anisotropy terms and obtain $P_{\mu^0}$, which is just the power spectrum that would be observed in the absence of redshift space distortions.

Going to longer observing times, it becomes possible to study the evolution of the anisotropy more quantitatively. While isolating the $\mu^4$ term---as was suggested by \cite{barkana2005}---seems unrealistic for the noise levels obtainable by LOFAR, the sum of the anisotropy terms can be used to extract information about the reionization history. We have shown that an inside-out reionization scenario gives an anisotropy that is initially positive, and later decreases and turns negative at around 20--30 per cent global ionized fraction for the range of $k$ modes considered here. While this anisotropy evolution alone may not be enough to distinguish the details of a particular reionization history, it provides an additional observable that reionization models will have to reproduce and can be used to exclude at least the more extreme outside-in models.

In conclusion, the subject of 21-cm redshift space distortions seems to warrant further attention. Far from being just a nuisance when interpreting observations, redshift space distortions are yet another example of the wealth of astrophysical and cosmological information that lies hidden in the 21-cm signal from the EoR. Among the most pressing questions in a short time-perspective is the universality of the anisotropy evolution that we have shown here. While it seems clear that extreme models, such as our outside-in toy model, can be excluded by LOFAR observations, it is not obvious how more subtle changes in the model assumptions will affect the anisotropy. 

A related issue is our assumption that $T_S \gg T_{\mathrm{CMB}}$. For this to be true, the spin temperature must be coupled to the gas temperature, and the IGM must be heated quickly by the first sources, before reionization gets started. If the heating phase is more extended, and overlaps with the reionization phase, the 21-cm signal will contain additional power from $T_S$ fluctuations \citep{ciardi2007,thomas2011}. Recently, \cite{mesinger2013} showed that in certain models, the spherically averaged 21-cm power spectrum can be enhanced by a factor 10--100 by spin temperature fluctuations in the early stages of reionization. If so, the 21-cm signal will be easier to detect, but since the $T_S$ fluctuations are likely correlated with the $x_i$ fluctuations, the quasi-linear approximation used here would no longer be valid, and the physical interpretation of the redshift space distortions would be more complicated. However, it may be possible to determine from observations whether a certain redshift lies in the high $T_S$ regime or not \citep{santos2008}.

It is also possible that the anisotropy evolution can better be extracted by assuming that the cosmological model is well known (eliminating the uncertainties in the $\mu^4$ term), or by using a different decomposition of the power spectrum. The $\mu$ decomposition used here has the advantage of offering simple physical interpretations of the different moments, but has the disadvantage of not forming an orthogonal basis---hence our focus on the sum of the anisotropy terms. Other decompositions, such as Legendre polynomials, avoid these problems, but the results may be more difficult to interpret. 

On longer time-scales, the Square Kilometre Array (SKA) will deliver observations with a signal-to-noise that will far exceed that of LOFAR, which may facilitate the extraction of the pure cosmological information contained in the $\mu^4$ term. However, in this low-noise regime, it becomes critical to understand the possible biases introduced by the foreground removal, which we have only touched upon in this paper.

%%%%%%%%%%%%%%%%%%%%%%%%%%%%%%%%%%%%%%%%%%%%%%%%

\section*{Acknowledgements}
This study was supported 
by the Swedish Research Council grants 2012-4144 and 2009-4088, the Science and
Technology Facilities Council [grant number ST/I000976/1]
and The Southeast Physics Network (SEPNet).
The authors acknowledge the Swedish National Infrastructure for Computing (SNIC)
resources at HPC2N (Ume\aa, Sweden) and PDC (Stockholm, Sweden) and
the Texas Advanced Computing Center (TACC) at The University of Texas at
Austin (\url{http://www.tacc.utexas.edu}) for providing HPC resources. This research
was supported in part by NSF grant AST-1009799, NASA grants NNX07AH09G and NNX11AE09G and TeraGrid grant TG-AST0900005.

FBA acknowledges the support of the Royal Society via a RSURF.

KKD is grateful for financial support from Swedish Research Council (VR) through the Oscar Klein Centre (grant 2007-8709). KKD would also like to thank the Indian Institute of Science and Educational Research, Kolkata and Center for Theoretical Studies, IIT Kharagpur for the hospitality they provided during the period when a part of this work has been done.

YM was supported by French state funds managed by the ANR within the Investissements d'Avenir programme under reference ANR-11-IDEX-0004-02.

MGS acknowledges support from FCT-Portugal under grant PTDC/FIS/100170/2008.

We would like to thank Matthew McQuinn for helpful discussions regarding the telescope noise simulations.
\footnotesize{
\bibliographystyle{mn2e} \bibliography{refs}
}

\label{lastpage}
\end{document}